\documentclass[sigconf]{acmart}
\usepackage{amsmath,amsfonts}
\usepackage[boxed]{algorithm}
\usepackage[noend]{algpseudocode}
\usepackage{array}
\usepackage{textcomp}
\usepackage{stfloats}
\usepackage{url}
\usepackage{soul}
\usepackage{verbatim}
\usepackage{graphicx}
\usepackage{wrapfig}
\usepackage{subcaption}
\usepackage{multirow}
\usepackage{paralist}
\newtheorem{Definition}{Definition}
\newtheorem{theorem}{Theorem}
\newtheorem{lemma}{Lemma}
\newtheorem{remark}{Remark}
\usepackage{xcolor}
\newcommand{\sd}[1]{{\color{red}{SD:  #1}}}
\newcommand{\ik}[1]{{\color{blue}{IK:  #1}}}

\title{Exploring The Resilience of Control Execution Skips against False Data Injection Attacks} 

\author{Ipsita Koley}
\affiliation{\institution{Indian Institute of Technology, Kharagpur}
	\country{India}}
\email{ipsitakoley@iitkgp.ac.in}

\author{Sunandan Adhikary}
\affiliation{\institution{Indian Institute of Technology, Kharagpur}
\country{India}}
\email{mesunandan@kgpian.iitkgp.ac.in}

\author{Soumyajit Dey}
\affiliation{\institution{Indian Institute of Technology, Kharagpur}
	\country{India}}
\email{soumya@cse.iitkgp.ac.in}

\setcopyright{none}
\renewcommand\footnotetextcopyrightpermission[1]{}

\acmYear{2022}

\begin{document}
\begin{abstract} 
Modern Cyber-Physical Systems (CPSs) are often designed as networked, software-based controller implementations which have been found to be vulnerable to network-level and physical level attacks. A number of research works have proposed CPS-specific attack detection schemes as well as techniques for attack resilient controller design. However, such schemes also incur platform-level overheads.  In this regard, some recent works have leveraged the use of skips in control execution to enhance the resilience of a CPS against false data injection (FDI) attacks. 
However, skipping the control executions may degrade the performance of the controller. 
		
In this paper, we provide an analytical discussion on when and how skipping a control execution can improve the system’s resilience against FDI attacks while maintaining the control performance requirement. We also propose a methodology to synthesize such optimal control execution patterns.  To the best of our knowledge, no previous work has provided any quantitative analysis about the trade-off between attack resilience and control performance for such aperiodic control execution. Finally, we evaluate the proposed method on several safety-critical CPS benchmarks.
\end{abstract}


\maketitle
\section{Introduction}
	\label{secIntroduction}
	Deployment of network components in  cyber-physical systems (CPSs) along with software-based, sophisticated control implementations have found wide applicability ranging from industrial control, connected-mobility to defense installations. However, such advancements have opened up different possible attack surfaces leading to network-level as well as physical-level attacks. Numerous  such attacks on safety-critical CPSs have been reported in the past, for example, Stuxnet\cite{langner2013kill}, Maroochy water breach attack\cite{slay2007lessons}, Black energy attack\cite{nazario2007blackenergy}, attacks in automotive domain\cite{greenberg2015hackers,checkoway2011comprehensive}, etc.
	\par In this work, we consider a type of attack called false data injection (FDI). Networked control CPSs are designed as a closed-loop where the controller receives the measurements from the plant,  computes the control signal such that the plant operates at/near the  desired reference point and sends the control signal to the plant. Sometimes, all the states of the plant can not be measured and  an observer (like Kalman filter) is used at the controller end to estimate the states of the plant. As the closed loop communication   happens over a network, an external or internal attacker can malign the sensor measurements and/or the control signals physically or through the network. In such FDI attacks, the controller and the plant do not receive actual data, leading to instability or performance loss.   

	\par Till date the best defence mechanism against FDI attacks is to use cryptographic methods. 
	However, the major hindrance to their application is the computation and communication loads incurred by these methods \cite{munir2018design,lesi2017security}. An alternate solution that can be found in literature is to use residue-based light-weight attack detection methods \cite{mo2010false, teixeira2015secure, koley2020formal} interleaved with traditional cryptographic techniques instead of using the latter continuously \cite{jovanov2019relaxing,adhikary2020skip}. 
 	A typical FDI attack can not maintain its stealthiness while the transmitted data is secured with cryptographic methods. On the other hand, the statistical nature of the residue-based detectors take some time to detect an FDI attack with higher probability. 
 	A smart attacker can intelligently craft the worst-possible FDI attack that can bypass the residue-based attack detectors when cryptographic methods are not active
 	\cite{teixeira2015secure,adhikary2020skip}. Therefore, irrespective of what security enforcement is in place (whether the combination of cryptographic method and residue-based detection or continuous use of cryptographic method), the FDI attacks 
 	can significantly affect the system's performance. 
Thus, the question that rises in this context is \emph{how to make the system more resilient against FDI attacks?} In this paper, our objective is to address this question.
	\par The authors of \cite{adhikary2020skip} were the first to explore the skipping of some control executions to enhance system's resilience against stealthy FDI attacks while it is not detected. Pattern-based execution of controller where some of the instances of control executions are dropped or skipped was initially studied to accommodate multiple tasks on resource constrained embedded platforms \cite{ghosh2017structured, majumdar2011performance,soudbakhsh2013co}. Skipping or dropping a control execution at a certain sampling instance means no new control input will be computed or communicated at that instance. So, the processor and the communication channel between the plant and controller both will also be free during that sampling instance. It is evident that the FDIs during skips are rendered ineffective. Thus, this can restrain the effectiveness of attacker's effort by enhancing system's resilience. But this skipping of control execution may degrade the performance of the control system. To address this, a \emph{minimum rate of control execution} is required to maintain the desired control performance \cite{ghosh2017structured}. Constraining this minimum rate of control execution, the authors of \cite{adhikary2020skip} developed a formal methodology-based approach to synthesize control execution-skipping sequences to enhance system security and safety. 
 However, the major limitations of their approach is that, it does not relate the position of control skips with the dynamics of the system under attack, and the SMT-based attack and control execution pattern synthesis might not scale for systems with larger dimension. Also,  if the position of control execution skips are not chosen wisely, the performance of the controller may degrade. 

Similar to \cite{adhikary2020skip}, we also utilize the skipping of control executions to enhance the resilience of the system under attack which may seem counter-intuitive as control performance may degrade due to execution skips.  
	However, the safety-critical CPSs mostly operate at higher sampling rate. The desired performance of such fast systems can be maintained if we can judiciously choose when to skip the control executions. Such control execution skips in turn will ignore that attacks injected at those sampling instances.
\par In this work, we theoretically analyse and establish analytical conditions under which execution skips will surely be beneficial in terms of enhancing the system's resilience against FDI attacks. We  present a methodology to synthesize the most optimal attack-resilient control execution sequence that also  ensures the desired performance. We provide an automated CAD tool-chain to generate such control execution sequences given any CPS. To this end, we now summarize the contributions of this work as follows.
	\begin{compactenum}
	    \item Given the specifications of a safety-critical CPS and its initial region, we formulate a constraint solving problem to generate the optimal or worst case FDI attack sequence that consumes minimum time to make the system unsafe.
		\item 
		We theoretically derive under which criteria the control execution skips will actually be favourable in enhancing system's resilience against FDI attacks.
		\item Utilizing the conditions established in previous contribution,
		we design a dynamic programming (DP) based solution methodology to synthesize the control execution patterns that ensure desired control performance as well as the best possible attack-resilience against optimal FDI attacks generated in contribution $1$ . 
		\item 
		We provide an automated CAD tool that takes as input the CPS   specification and synthesizes resilient control execution patterns for the same. The scalability of the proposed methodology has been   evaluated on well known benchmarks with  various dimensions.
		
	\end{compactenum}

	\begin{figure}[!hb]
		\centering
		\includegraphics[scale=0.25,clip]{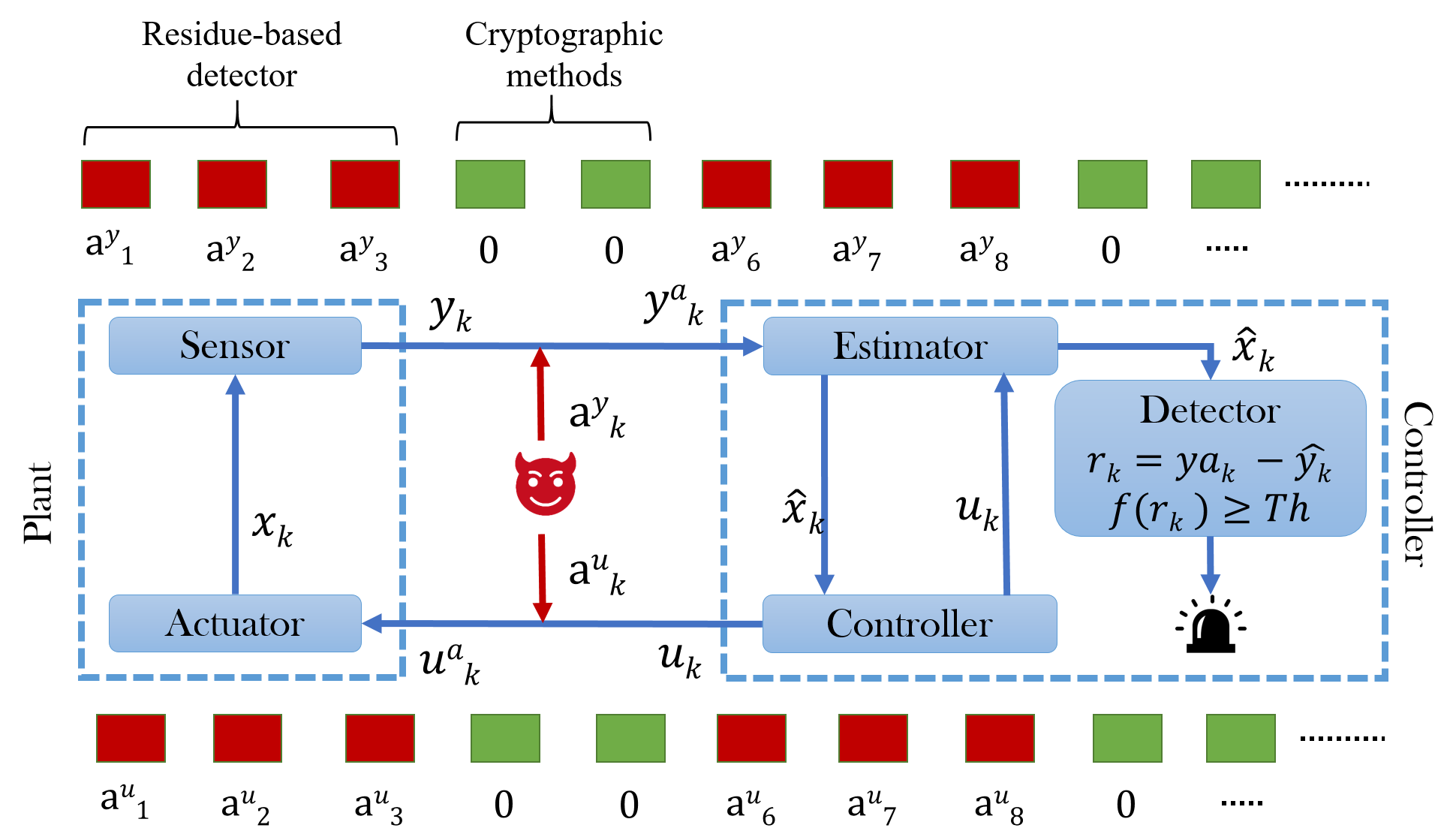}\vspace*{-2mm}
		\caption{Secure CPS architecture}
		\label{figCPSAttackModel}
	\end{figure}
	
	\section{Background}
	\label{secBackground}
	\textbf{Secure CPS Model: } 
	General architecture of a secure CPS is presented in Fig.~\ref{figCPSAttackModel}. The physical process i.e. the plant and the controller work together in a closed loop manner such that the desired operating criteria of the physical process is maintained. They communicate between themselves over a network, which we consider is vulnerable to FDI attacks. In the absence of an adversary, the closed loop dynamics of a CPS can be presented as a discrete linear time-invariant (LTI) system like the following.
	
	{\footnotesize
	\begin{align}
		\nonumber
		x_{k+1} &= Ax_k + Bu_k + w_k;\ y_k = Cx_k + v_k;\ 	\hat{y}_{k+1} = C(A\hat{x}_k + Bu_k);\\ 
	 r_{k+1} &= y_{k+1} - \hat{y}_{k+1};\
		\hat{x}_{k+1} = A\hat{x}_k + Bu_k + Lr_{k+1};\ u_k = -K\hat{x}_k;\ e_{k} = x_k - \hat{x}_k;
		\label{eqLTINoAttack}
	\end{align}	
	}
	Here, $x_k \in \mathbb{R}^n$ is the system state vector,   $y_k \in \mathbb{R}^m$ is the measurement vector obtained from available sensors at $k$-th time stamp; $A, B, C$ are the system matrices. We consider that the initial state $x_0 \in \mathcal{N}(\bar{x}_0, \Sigma)$, the process noise $w_{k}\in \mathbb{R}^n \sim \mathcal{N}(0, \Sigma_w)$ and the measurement noise $v_{k} \in \mathbb{R}^m \sim \mathcal{N}(0, \Sigma_v)$ are independent Gaussian random variables. Further, in every $k$-th sampling instant, the observable system state $\hat{x}_k$ is estimated using system output $y_k$ while minimizing the effect of noise, and used for computing the control input $u_k \in \mathbb{R}^l$. The estimation error $e_k$ is defined as the difference between actual system states $x_k$ and estimated system states $\hat{x}_k$. We denote the residue i.e. the difference  between the measured and the estimated outputs as $r_k$. The estimator gain $L$ and controller gain $K$ are designed in such a way that it is ensured both $(A-LC)$ and $(A-BK)$ are stable. As security enforcement, we consider a sporadic implementation of some cryptographic method along with a residue-based detector as demonstrated in Fig.~\ref{figCPSAttackModel}. The residue-based detector computes a function $f(r_k)$ ($f$ can be a simple norm or any statistical method, like $\chi^2$-test) and compares it with a threshold $Th$ to identify any anomalous behavior of the system. 
	\par Consider an FDI attack, where the attacker injects false data $a^y_k$ and $ a^u_k$ (Fig.~\ref{figCPSAttackModel}) to the sensor measurement and control signal respectively when the cryptographic method is not active (see Fig.~\ref{figCPSAttackModel}). In such scenario, the system dynamical equation becomes,
	
	{\footnotesize
	\begin{align}
		\nonumber
		x^a_{k+1} &= Ax^a_k + B\tilde{u}^a_k + w_k;\ y^a_k = Cx^a_k + v_k + a^y_k\\ 
		\nonumber
		\hat{y}^a_{k+1} &= C(A\hat{x}^a_k + Bu^a_k);\ 
		r^a_{k+1} = y^a_{k+1} - \hat{y}^a_{k+1}\\ 
		\hat{x}^a_{k+1} &= A\hat{x}^a_k + Bu^a_k + Lr^a_{k+1};
		u^a_k = -K\hat{x}^a_k;\ \tilde{u}^a_k = u^a_k + a^u_k;\ e^a_k = x^a_k - \hat{x}^a_k;
		\label{eqLTIUnderAttack}
	\end{align}
	}
	Here, $x^a_k$, $\hat{x}^a_k$, $y^a_k$, $r^a_k$, $u^a_k$, $\tilde{u}^a_k$, and $e^a_k$ represent plant state, estimated plant  state, forged sensor data, residue, control signal, forged control signal, and estimation error respectively in an FDI attack scenario. $u^a_k$ is the control input computed at $k$-th sampling instance on which the effect previous attacks i.e. $a^u_i$ and $a^y_i$ for $1\leq i\leq k-1$ persist. 	$u^a_k$ added with attack on actuator $a^u_k$ at $k$-th sample produces $\tilde{u}^a_k$ i.e. the forged control signal at $k$-th sample. Note that even though we discussed about intrusion through network, physical level sensor data tampering \cite{shoukry2013non} can also happen. The above attack model is generic to all kind of such falsification attacks. We denote an attack vector at $k$-th sampling instance as $\mathcal{A}[k]= [a^u_k,a^y_k]^T$. If the attacker continues the false data injection for $l$ sampling iterations, then the $l$ length attack vector is expressed as follows $\mathcal{A}_l = [\mathcal{A}[1]\cdots\mathcal{A}[l]]=\begin{bmatrix}
 a^u_1 & \cdots &  a^u_l\\ 
 a^y_1 & \cdots &  a^y_l
\end{bmatrix}$. Falsifying the control input by injecting a sequence of $a^u$'s, the attacker forces the states of the system to go beyond the safety limit. On the other hand, it modifies the sensor measurements with a sequence of $a^y$'s such that it can hide itself from the residue based detector. We define such \emph{stealthy} FDI attack vector as follows.
\vspace{-0.2mm}
	\begin{Definition}[\bf Stealthy false data injection attack] An attack vector $\mathcal{A}_l = [\mathcal{A}[1]\cdots\mathcal{A}[l]]==\begin{bmatrix}
 a^u_1 & \cdots &  a^u_l\\ 
 a^y_1 & \cdots &  a^y_l
\end{bmatrix}$ of length $l$ is said to be stealthy if $f(r^a_k) < Th\ \forall k\in[1,l]$ where $r^a_k$ is the residue generated due the attack vector $\mathcal{A}_l$ at $k$-th sampling instance. \hfill$\Box$
\label{defStealthy}
	\end{Definition}
	
\par\noindent	\textbf{Control Execution Skip Pattern: }
	When we say that a control execution is skipped in a certain $k$-th sampling instance, the implication of the same on the underlying system are as follows.
	\begin{compactenum}
		\item The sensor measurements $y_k$ are not  communicated to the controller unit. 
		\item A fresh control input $u_k$ is not  calculated and communicated to the plant. The plant updates its states simply using the  previous control input
		\item Detection unit will also not operate.
	\end{compactenum}
	We consider that such skips in control execution shall be regular leading to a pattern in lines of \cite{ghosh2017structured}. This is naturally required for deterministic system design and deployment. 
	We provide a formal definition of control execution skipping pattern.
	\begin{figure}[!ht]
		\centering
	\includegraphics[width=\columnwidth]{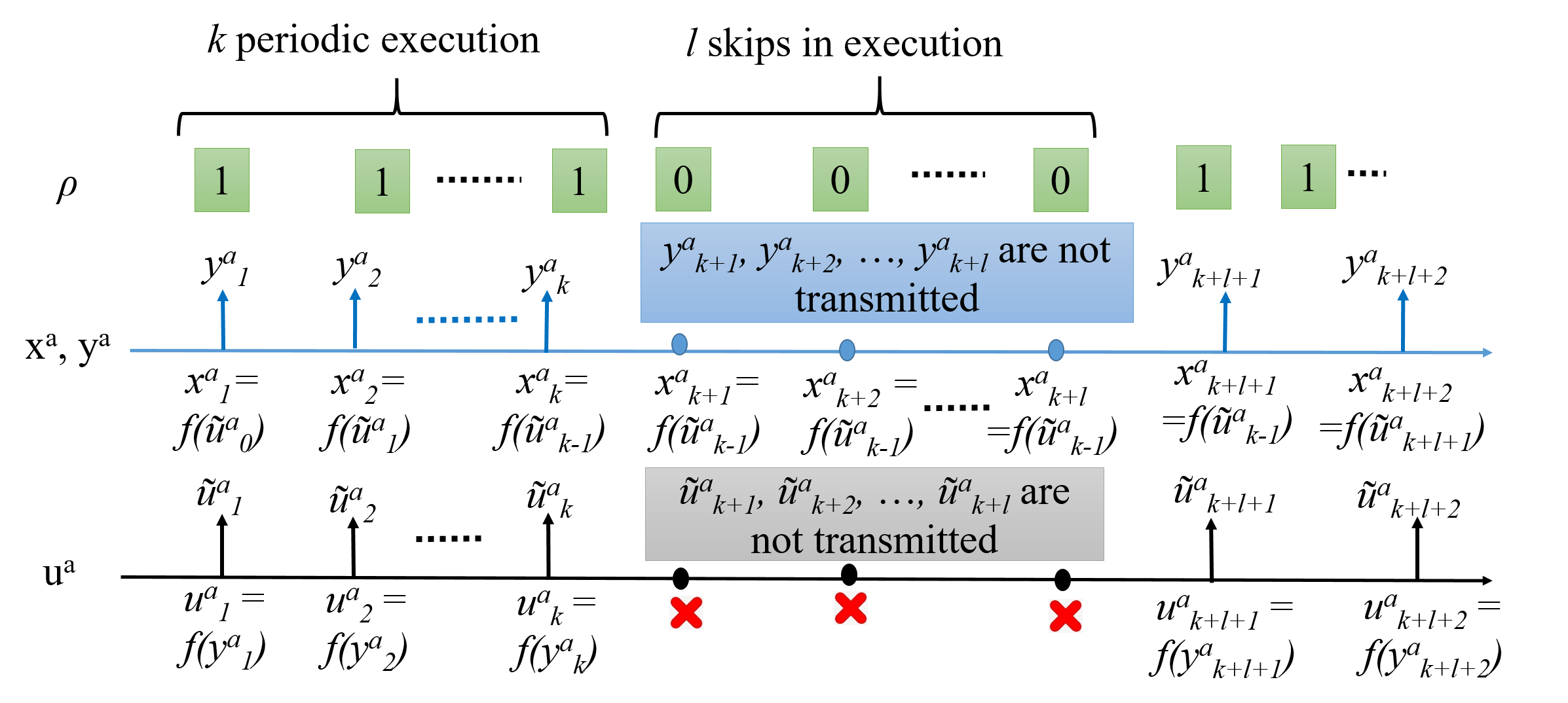}\vspace*{-2mm}
	\caption{Implication of control skip pattern on the system}
	\label{figPattern}
	\end{figure}
	\begin{Definition}[\bf Control Execution Skip Pattern]
		A $t$ length control execution skip pattern for a given control loop $(A,B,C,K,L)$, is a sequence $\rho\in \{0, 1 \}^t$ such that it can be used to define an infinite length control execution sequence $\pi$, repeating with period $t$, defined as,  $\pi[k]=\pi[k+t] = \rho[k\%t], \forall k\in \mathbb{Z}^+$ where $A$, $B$, and $C$ are system matrices, $K$ is the controller gain, and $L$ is the observer gain of the system. \cite{ghosh2017structured}. \hfill$\Box$
	\end{Definition}
	\noindent Symbolically, a pattern can be denoted as $(1^k0^l)^t$ where $k, t > 0$ and $l\geq 0$. In a pattern $\rho$, $1$ denotes control execution and $0$ denotes control execution skip. Some examples of patterns are $1^t = $ (periodic execution i.e. $0$ skip), $(10)^t = 101010\cdots$, $111001010\cdots$, etc. Consider that there is a skip in control execution at $(k+1)$-th sample as demonstrated in Fig.~\ref{figPattern}. Then according to skip properties, $u_k = u_{k-1}$, $a^u_{k} = a^u_{k-1}$, and $\triangle r_{k+1} = 0$. Therefore, plant state $x_{k+1}$ at $(k+1)$-th sample will be updated by old control input i.e. $u_{k-1}$. The notion of skip changes the system dynamics like the following.
	\vspace{-2mm}
	
	{\footnotesize
	\begin{align}
		\nonumber
		x^a_{k+1} &= Ax^a_k + B(u^a_{k-1} + a^u_{k-1}) + w_k;\ \hat{x}^a_{k+1} = A\hat{x}^a_k + Bu^a_{k-1};\\
		\nonumber
		x_{k+1} &= Ax_k + Bu_{k-1} + w_k;\ \hat{x}_{k+1} = A\hat{x}_k + Bu_{k-1};\\
		e^a_{k+1} &= Ae^a_k + Ba^u_{k-1} + w_k;\ e_{k+1} = Ae_k + w_k;
		\label{eqStateInSkip}
	\end{align}}
	\subsubsection*{Control Performance:} In this work, we define the control performance with respect to the settling time $T_s$ of a system. Settling time is the duration within which the system output must reach and stay within $2\%$ band of the reference. To ensure the desired performance while some of the control executions are skipped, the control execution must maintain the minimum rate $r_{min}$ \cite{ghosh2017structured} such that the settling time property can be achieved. For example, in a window of $t$ samples, the controller must be executed $\lceil t\times r_{min}\rceil$ times. This implies in a control execution skip pattern $\rho$ of length $t$, there must be at least $\lceil t\times r_{min}\rceil$ 1's.
	
	
	\section{Problem Formulation}
	\label{secProbForm}
\begin{figure}[!h]
		\captionsetup{justification=centering}
		\centering
		\begin{subfigure}[b]{.5\columnwidth}
			\centering
			\includegraphics[width=\textwidth,keepaspectratio,clip]{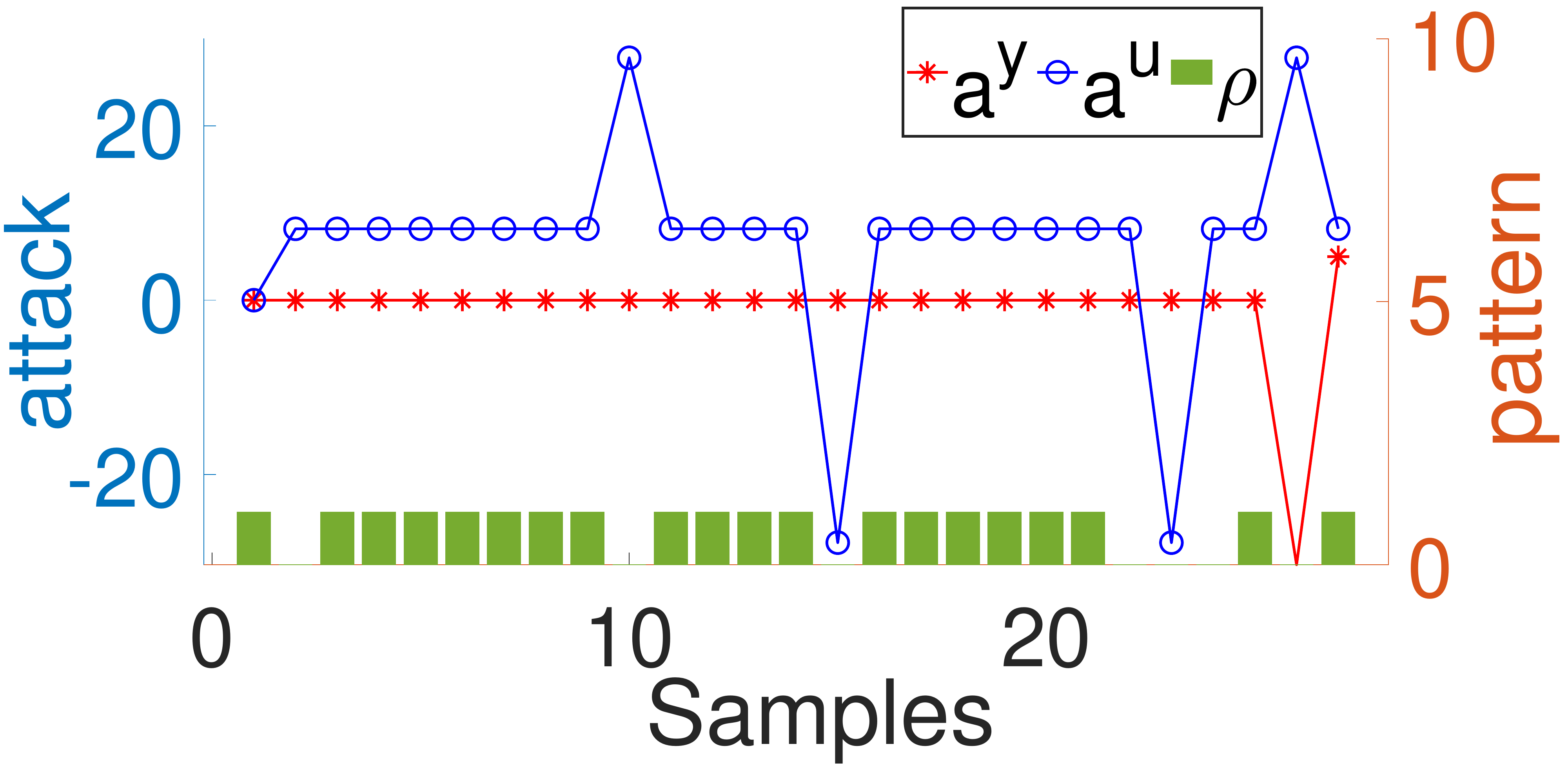}
			\caption{Attack vector and control skip pattern}
			\label{figAttackPatternMot}
		\end{subfigure}
		\hfill
		\begin{subfigure}[b]{.48\columnwidth}
			\centering
			\includegraphics[width=\textwidth,keepaspectratio,clip]{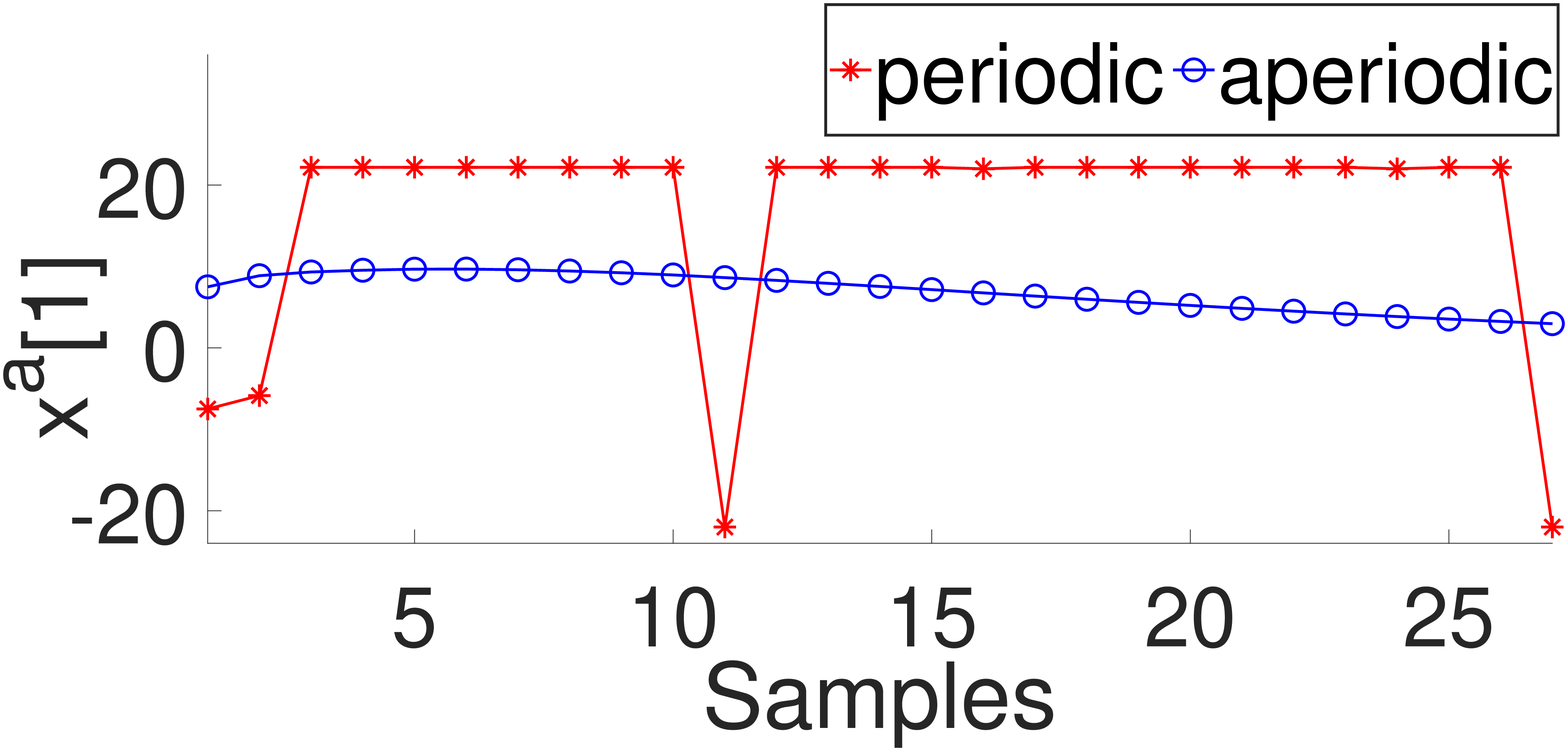}
			\caption{Effect of pattern on state 1}
			\label{figState1Mot}
		\end{subfigure}
		\begin{subfigure}[b]{.5\columnwidth}
			\centering
			\includegraphics[width=\textwidth,keepaspectratio,clip]{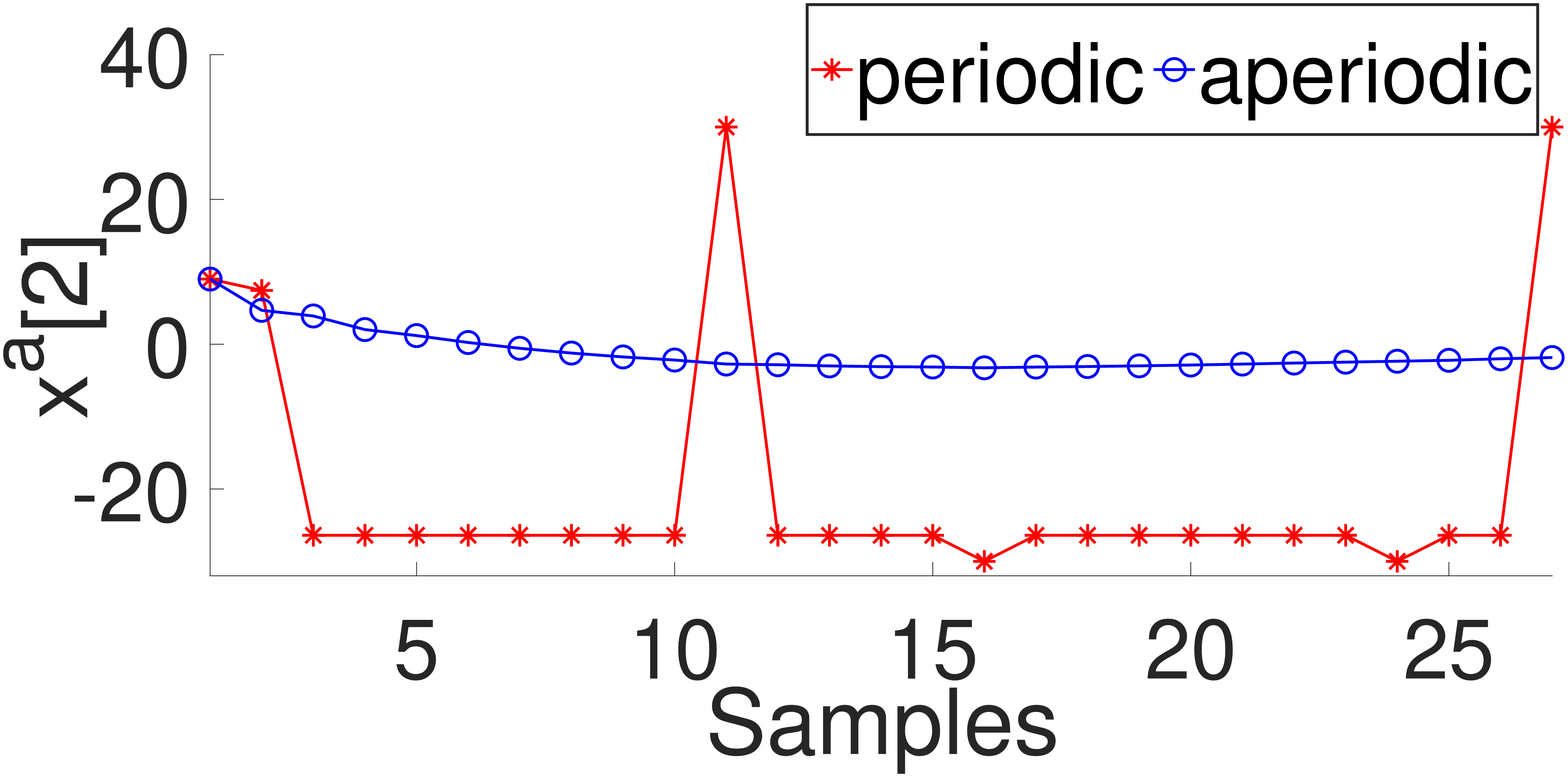}
			\caption{Effect of pattern on state 2}
			\label{figState2Mot}
		\end{subfigure}%
		\hfill
		\begin{subfigure}[b]{.5\columnwidth}
			\centering
			\includegraphics[width=\textwidth,keepaspectratio,clip]{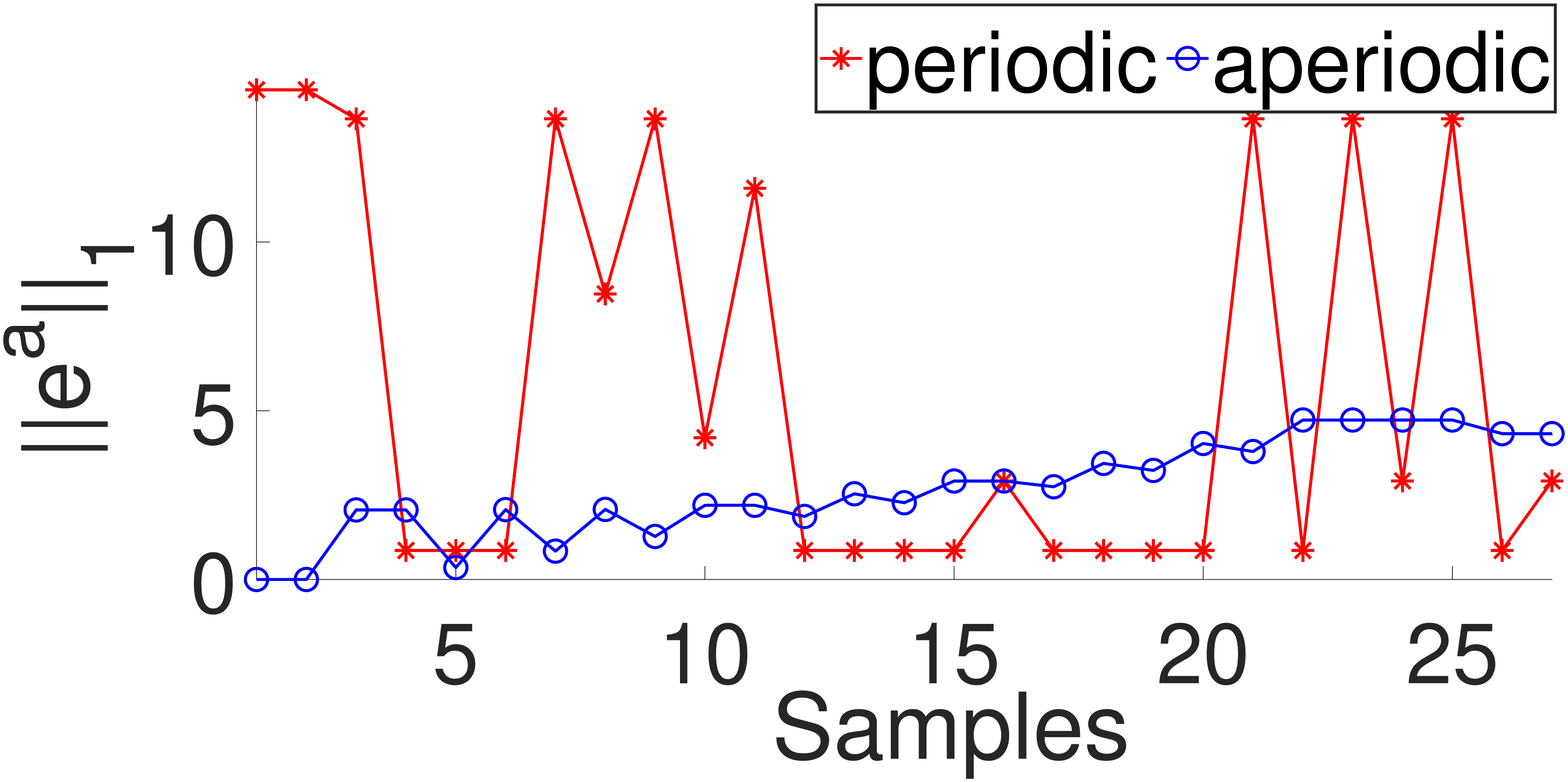}
			\caption{Effect of pattern on estimation error}
			\label{figEstErrorMot}
		\end{subfigure}
		\caption{Demonstrating  the effect of control skip pattern on system's resilience against FDI attacks}
		\label{figMotExample}
	\end{figure}

	\textbf{A Motivating example: } In this section, we demonstrate how occasional skips in control execution improve the system's resilience against FDI attacks with help of a motivating example (Fig.~\ref{figMotExample}). We take an example of trajectory tracking control (TTC) system of a vehicle from \cite{adhikary2020skip}. This is a 2-dimensional system with the $deviation$ from the reference trajectory and $velocity$ of the vehicle as states. Attacker adds false data to the measurement data(i.e. deviation) and the control signal(i.e. acceleration).
	In Fig.~\ref{figAttackPatternMot}, the attacks on measurement $a^y$ and control signal $a^u$ are given. We design a control skip pattern $\rho$ by introducing drops in control executions with an intention to weaken the attack's effect. Thus, we inject the skip where the attack on actuation $||a^u||$ increases (green bar graph in Fig.~\ref{figAttackPatternMot}). The minimum execution rate of the controller is also more than the requirement i.e. $50\%$ \cite{adhikary2020skip}. The effect of introducing drops in the presence of the attacker is presented in Fig.~\ref{figState1Mot}-\ref{figEstErrorMot}. We can observe that when the attacker injects optimal attack values (Fig.~\ref{figAttackPatternMot}), there is significant deviation in system's state progression (Fig.~\ref{figState1Mot} and \ref{figState2Mot}) when periodic control execution takes place. However, due to introduction of drops in the control execution, FDIs rendered ineffective on system's state progression (Fig.~\ref{figState1Mot} and \ref{figState2Mot}) in case of aperiodic control execution. This is because the FDIs in periodic execution increases the estimation error $e^a$ considerably (Fig.~\ref{figEstErrorMot}). This in turn affects the control performance poorly (According to Eq.~\ref{eqLTIUnderAttack}). On the other hand, ignorance of the attack values on actuation signal during drops minimizes the attack's effect on $e^a$, thereby contains $e^a$ within much lower range (Fig.~\ref{figEstErrorMot}). Thus, we can see the aperiodic control execution following the pattern $\rho$( Fig.~\ref{figAttackPatternMot}) enhances TTC's resilience against the FDI attack given in Fig.~\ref{figAttackPatternMot}. Keeping in mind the stealthy FDI attack and its effect on system's state progression, how to generate the pattern to weaken the attack effect most effectively, is discussed in Sec.~\ref{secProposedMethod}. In the following section, we provide a quantitative analysis behind such motivation of this work.
	
	\par\noindent\textbf{Analysing effect of execution skips : } 
	During execution skips, the attacks on sensor data and control signal are ignored. This seems useful with respect to attack resiliency. But, during control execution skips no new control input is computed and communicated to the plant. The plant updates its states using the previous control input. This may lead to poor control performance.  Naturally, there exists a trade-off between resilience against attack and the amount of  control performance to forego during skips. Our aim is to provide a formal discussion on when and how skips can improve the resilience against FDI attacks and at the same time desirable control performance is maintained when the control execution is aperiodic. First, to capture the difference in system's response in presence and absence of FDI attacks in case of periodic control execution, we introduce the following two terms:
	\vspace{-2mm}
	
	{\footnotesize
	\begin{align}
		\nonumber
		\triangle e_k &= e^a_k - e_k\ = (A-LCA)\triangle e_k + (B - LCB)a^u_k - La^y_{k+1}\\
				&= A\triangle e_k + Ba^u_k - L\triangle r_{k+1}\ = \sum_{i=0}^{k-1} A^i(Ba^u_{k-1-i} - L\triangle r_{k-i})\ (\triangle e_0 = a^y_0 = 0) \label{eqEstErrDeviationPeriodic}\\
				\triangle r_k &= r^a_k - r_k = CA\triangle e_k + CBa^u_k + a^y_{k+1}\label{eqResDeviationPeriodic}
		\end{align}
		}
	Here, $\triangle e$ and $\triangle r$ present how much the estimation error and residue vary due to the FDI attack. Let us consider, there is a skip in control execution at $(k+1)$-th sample. Then following Eq.~\ref{eqStateInSkip} we get,
	
	{\footnotesize
	\begin{align}
		\nonumber
		\triangle e_{k+1} &= e^a_{k+1} - e_{k+1}\ = (Ae^a_k + Ba^u_{k-1} + w_k) - (Ae_k + w_k);\\
		&= A\triangle e_k + Ba^u_{k-1}\label{eqEstErrorAperiodic}\\
		\triangle r_{k+1} &= 0		
	\end{align}}
	
	Now, to show whether the execution skips actually enhance the resilience of the system against FDI attacks, we compare two parameters: i) $\triangle e^p_{k+l}$ i.e. estimation error deviation after $(k+l)$ periodic executions and ii) $\triangle e^{ap}_{k+l}$ i.e. estimation error deviation after $k$ periodic executions followed by $l$ control execution drops. This is demonstrated in Fig.~\ref{figPattern}. The control execution is periodic during 1st $k$ samples. The closed loop system progresses following Eq.~\ref{eqLTIUnderAttack}. At $(k+1)$-th sample when the control execution is dropped, measurement of that sampling instance $y^a_k$ is not transmitted, the control input $u^a_k$ is neither computed  and transmitted, and the state progresses with last received control signal $\tilde{u}^a_k$ (Eq.~\ref{eqStateInSkip}). This is continued till $(k+l)$-th sample. Once, the controller receives new measurement at $(k+l+1)$-th sample, the sensor measurement $y^a_{k+l+1}$ is transmitted to the plant, the control input $u^a_{k+l+1}$ is again computed using $y^a_{k+l+1}$ (Eq.~\ref{eqLTIUnderAttack}) and transmitted to the plant. The plant updates its states following Eq.~\ref{eqLTIUnderAttack} using new control input.
	Now, from Eqs.~\ref{eqEstErrDeviationPeriodic} and \ref{eqEstErrorAperiodic}, we capture the iterative forms of $\triangle e^p_{k+l}$ and $\triangle e^{ap}_{k+l}$ as,
	
	{\footnotesize
	\begin{align}
		\nonumber
		\triangle e^p_{k+l} &= A\triangle e_{k+l-1} + Ba^u_{k+l-1} - L\triangle r_{k+l}\\
		\nonumber
		&=A[A\triangle e_{k+l-2}+Ba^u_{k+l-2}-L\triangle r_{k+l-1}] + Ba^u_{k+l-1} - L\triangle r_{k+l}\\
		\nonumber
		&= A^2\triangle e_{k+l-2} + (ABa^u_{k+l-2} + Ba^u_{k+l-1}) - (AL\triangle r_{k+l-1} + L\triangle r_{k+l})\\
		\nonumber
		&=\cdots\\
		\nonumber
		&= A^l\triangle e_k +(A^{l-1}Ba^u_k+A^{l-2}Ba^u_{k+1}+\cdots+Ba^u_{k+l-1})\\
		\nonumber
		&-(A^{l-1}L\triangle r_{k+1}+A^{l-2}L\triangle r_{k+2}+\cdots+L\triangle r_{k+l})\\
		&= A^l\triangle e_k + \sum_{i=0}^{l-1}A^i(Ba^u_{k+l-1-i}-L\triangle r_{k+l-i})\label{eqEstPeriodic}
		\end{align}}
		{\footnotesize
		\begin{align}
		\nonumber
		\triangle e_{k+l}^{ap} &= A\triangle e_{k+l-1} + Ba^u_{k-1} = A[A\triangle e_{k+l-2} + Ba^u_{k-1}] + Ba^u_{k-1}\\
		\nonumber
		&=A^2\triangle e_{k+l-2} + (A + I)Ba^u_{k-1}\\
		\nonumber
		&=\cdots\\
		&= A^l\triangle e_k + (A^{l-1} + A^{l-2} + \cdots + 1)Ba^u_{k-1} = A^l\triangle e_k + \sum_{i=0}^{l-1}A^iBa^u_{k-1}
		\label{eqEstAperiodic}
	\end{align}
	}
	The first terms in both Eq.~\ref{eqEstPeriodic} and \ref{eqEstAperiodic} represent estimation error deviation up to $k$-th sampling instance. In both periodic and aperiodic cases, this term will be same. And, the second term in Eq.~\ref{eqEstPeriodic} and \ref{eqEstAperiodic} captures effect of last $l$ iterations in periodic and aperiodic cases respectively. We define the \emph{resilience} of a system against FDI attacks with respect to the terms $\triangle e^{p}$ and $\triangle e^{ap}$. If the FDI attack fails to do much harm to the system, the values of $\triangle e^{p}$ and $\triangle e^{ap}$ will be less. Thus, lower values of $\triangle e^{p}$ and $\triangle e^{ap}$ imply that the system is more resilient against the FDI attacks. Occasional skips in control execution will be useful in enhancing system's resilience against FDI attacks if the difference in estimation error in periodic execution due to attack $\triangle e^p$ is more than that of aperiodic control execution $\triangle e^{ap}$. In the following theorem we establish under which condition  $\triangle e^p$ will be more than $\triangle e^{ap}$.
	\vspace{-2mm}
	\begin{theorem}
	\label{thCriteria}
	For a plant-controller closed-loop system under FDI attack (Eq.~\ref{eqLTIUnderAttack}), control execution skips for consecutive $l$ sampling instances after $k$ periodic control executions will be effective in enhancing system's resilience when the following criteria is true:\\ $||\sum_{i=0}^{l-1}A^iB\triangle a^u_{k+l-1-i}||$ $> ||\sum_{i=0}^{l-1}L\triangle r_{k+l-i}||$\\
		Here, the term $\triangle a^u_{k+l-1-i} = a^u_{k+l-1-i}-a^u_{k-1}, \forall i \in [0,l-1]$ i.e. captures the difference between the attacks on control signal on the last $l$ sampling instances out of $(k+l)$ samples between periodic (i.e. none of the $k+l$ samples has been skipped) and aperiodic cases(i.e. $l$ out of the $k+l$ samples has been skipped).\hfill$\Box$
	\end{theorem}
	\emph{Proof:} The control execution skips for consecutive $l$ sampling instances after periodic execution of consecutive $k$ sampling instances will enhance system's resilience against FDI attacks if $||\triangle e^{p}_{k+l}|| > ||\triangle e^{ap}_{k+l}||$. Now,
	
	{\footnotesize
	\begin{align}
		\nonumber
		&||\triangle e^{p}_{k+l}|| > ||\triangle e^{ap}_{k+l}||\\
			\nonumber
		\implies &||A^l\triangle e_k + \sum_{i=0}^{l-1} A^i(Ba^u_{k+l-1-i}-L\triangle r_{k+l-i})|| >||A^l\triangle e_k + \sum_{i=0}^{l-1} A^iBa^u_{k-1}||\\
			\nonumber
		\implies &||A^l\triangle e_k|| + ||\sum_{i=0}^{l-1} A^i(Ba^u_{k+l-1-i}-L\triangle r_{k+l-i})||>||A^l\triangle e_k + \sum_{i=0}^{l-1} A^iBa^u_{k-1}||\\
		\nonumber
		\implies &||A^l\triangle e_k||-||A^l\triangle e_k + \sum_{i=0}^{l-1} A^iBa^u_{k-1}||>-||\sum_{i=0}^{l-1} A^i(Ba^u_{k+l-1-i}-L\triangle r_{k+l-i})||\\
		\nonumber
		\implies &||-\sum_{i=0}^{l-1} A^iBa^u_{k-1}||>||\sum_{i=0}^{l-1} A^i(L\triangle r_{k+l-i}-Ba^u_{k+l-1-i})||\\
		\nonumber
		\implies &||-\sum_{i=0}^{l-1} A^iBa^u_{k-1}|| >||\sum_{i=0}^{l-1} A^iL\triangle r_{k+l-i}||-||\sum_{i=0}^{l-1} A^iBa^u_{k+l-1-i}||\\
		\nonumber
		\implies &||\sum_{i=0}^{l-1} A^iBa^u_{k+l-1-i}|| - ||\sum_{i=0}^{l-1} A^iBa^u_{k-1}||>||\sum_{i=0}^{l-1} A^iL\triangle r_{k+l-i}||\\
		\nonumber
		\implies &||\sum_{i=0}^{l-1} A^iBa^u_{k+l-1-i} - \sum_{i=0}^{l-1} A^iBa^u_{k-1}||>||\sum_{i=0}^{l-1} A^iL\triangle r_{k+l-i}||\\
		\implies &||\sum_{i=0}^{l-1}A^iB\triangle a^u_{k+l-1-i}||>||\sum_{i=0}^{l-1} A^iL\triangle r_{k+l-i}||\quad\quad\quad\quad\quad\Box
		\label{eqEpMoreThanEap}
	\end{align}
}
	\begin{remark}
	\label{remNonEqAttack}
	For an aperiodic control execution pattern of $k$ consecutive periodic execution followed by $l$ control skips, the effect of actuation attack at $(k-1)$-th sample (starting from the $0$-th sample) i.e. $a^u_{k-1}$ gets forwarded through all of the next $(l-1)$ iterations. Whereas, in case of periodic control executions, an attacker can inject different actuation attack values $a^u_{k+l-1-i}$ at each of the sampling iterations, i.e., $\forall i\in[0,l-1]$. The above theorem states that whenever the attacker attempts to vary the attack efforts in consecutive iterations in order to stay stealthy or jeopardise the system safety faster (i.e., $||\sum_{i=0}^{l-1}A^iB\triangle a^u_{k+l-1-i}||\neq 0$), skipping the control execution at that sampling instance helps make the system more resilient against the injected false data.
\hfill$\Box$
	\end{remark}
	\subsubsection*{\textbf{Formal Problem Statement:}} Consider a plant with system matrices $A$, $B$, and $C$, controller $K$, observer $L$, and its initial region $X_0$. We now formally define the attack-resilient control execution skipping pattern as follows. \\
	\emph{For the given system specifications $\langle A, B, C, K, L, X_0 \rangle$, how we can find attack-resilient control execution skipping patterns $\rho = (1^k0^l)^t$ where $n,k>0$, $l\geq 0$ utilizing the relation $||\sum_{i=0}^{l-1}A^iB\triangle a^u_{k+l-1-i}||>||\sum_{i=0}^{l-1} A^iL\triangle r_{k+l-i}||$ provided minimum execution rate $r_{min}$ of the controller is maintained}?
	\section{Proposed Methodology}
	\label{secProposedMethod}
\begin{figure}[H]
	    \centering
	    \includegraphics[width=\columnwidth]{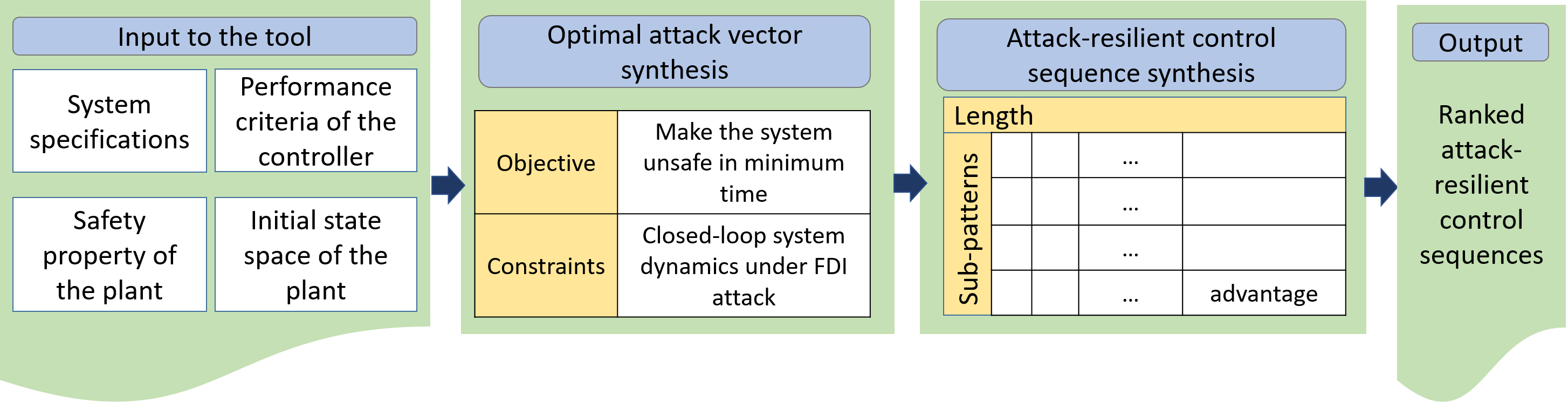}\vspace*{-3mm}
	    \caption{Framework for attack-resilient control sequence synthesis}
	    \label{figToolFlow}
	\end{figure}

	In the last section, we identified analytical  conditions, which if satisfied can make  skips in control execution to be beneficial in enhancing a system's resilience against stealthy FDI attacks. Now, we present our proposed framework for synthesizing attack-resilient control execution patterns. The outline of the framework is demonstrated in  Fig.~\ref{figToolFlow}. 
	The framework (Fig.~\ref{figToolFlow}) requires the followings inputs: i) System specifications matrices $A$, $B$, and $C$, controller gain $K$, observer gain $L$, the maximum limit $\mathcal{Y}$ of the sensor measurements, actuation saturation limit $\mathcal{U}$, and the threshold $Th$ of the detector in place, ii) Performance criteria of the controller i.e. minimum execution rate $r_{min}$, iii) The safety property of the system, defined as a safety polytope $X_s\in\mathbb{R}^n$ which mandates that the system trajectory will always be within $X_s$, and iv) Initial region of the plant states $X_0\in\mathbb{R}^n$ from which the system progression initiates. With these inputs, the framework sequentially runs two primary functional modules (Fig.~\ref{figToolFlow}). The \emph{first} one synthesizes an attack vector that consumes minimum time to make the system unsafe considering the system may initiate any where from $X_0$. The \emph{second} one generates attack-resilient control execution sequences based on the synthesized attack vector from the first module using a DP based method. 
	These two functional modules are elaborately discussed in Sec.~\ref{subsecAttackGeneration} and \ref{subsecPatternSyn} respectively.
	\subsection{Minimum-length Attack Vector Generation}
	\label{subsecAttackGeneration}
	For a given CPS, the notion of minimum length and stealthy false data injection attack is presented in the following definition.
	\begin{Definition}[\bf Minimum length Stealthy False Data Injection attack] A $t$ length false data injection attack vector $\mathcal{A}_t$ is stealthy (Def.~\ref{defStealthy}) and of minimum length if it can stealthily steer the system trajectory beyond the safety envelope $X_s$ while no attack vector of smaller length can make it possible i.e. $x^a_t\notin X_s$ and $f(r^a_k)<Th\ \forall k\in[1,t]$ but $x^a_k\in X_s\ \forall k\in[1, t-1]$.\hfill$\Box$
	\label{defMinLenAttack}
	\end{Definition}
	We formally present the problem of generating minimum length stealthy attack vector as follows.
	
	{\footnotesize
		\begin{align}
	    \mathcal{CP}: &\exists\ \mathcal{A}[1], \mathcal{A}[2], \cdots, \mathcal{A}[t]\quad \forall x\in X_0\label{eqMinLenAtt}\\
	\textrm{s.t.} \quad &x^a_0 = x;\ \hat{x}^a_0 = x\label{eqAttInit}\\
	\quad &u^a_{i-1} = -K\hat{x}^a_{i-1};\ \tilde{u}^a_{i-1} = u^a_{i-1} + a^u_i;\ \forall i\in[1,t]\label{eqAttControl}\\
	\quad &x_{i} = \mathcal{A}x_{i-1} + \mathcal{B}\tilde{u}^a_{i-1};\ y^a_i = Cx^a_i + a^y_i;\ \forall i\in [1,\ t]\label{eqAttState}\\
	& r^a_{i-1} = y^a_i - C(A\hat{x}^a_{i-1} + Bu^a_{i-1})\ \hat{x}_{i} = \mathcal{A}\hat{x}_{i-1} + \mathcal{B}u^a_{i-1} + Lr^a_{i-1}\ \forall i \in [1, t]\label{eqAttEst}\\
	& f(r^a_{i-1})<Th;\ |y^a_i|,|a^y_i|<\mathcal{Y};\ |u^a_i|, |\tilde{u}^a_i|, |a^u_i| < \mathcal{U}\ \forall i\in [1,\ t]\label{eqAttStealth}\\
	& x^a_i\in X_s \forall i\in [1,\ t-1];\ x^a_t\in X_s\label{eqAttSafety}
		\end{align}
	}
	The above constraint solving problem $\mathcal{CP}$ returns an attack vector $\mathcal{A}_t =$ $[\mathcal{A}[1]$, $\mathcal{A}[2]$, $\cdots$, $\mathcal{A}[t]]$ (\ref{eqMinLenAtt}) of length $t$ satisfying all the constraints in (\ref{eqAttInit})-(\ref{eqAttInit}) for any initial value of the state (\ref{eqMinLenAtt}). The constraints (\ref{eqAttControl})-(\ref{eqAttEst}) follow the system progression under attack (Eq.~\ref{eqLTIUnderAttack}). The stealthiness of $\mathcal{A}_t$ is ensured by the constraint $f(r^a_{i-1})<Th$ in \ref{eqAttStealth}.	The other constraints in (\ref{eqAttStealth}) guarantee that the attack is stealthy and attacks on sensor and actuation signal as well as the falsified measurement and actuation signals are within their respective ranges. To make sure the attack vector $\mathcal{A}_t$ is of minimum length (Def.~\ref{defMinLenAttack}), we keep the safety constraints in (\ref{eqAttSafety}). Initially, we solve this problem using some constraint solver with value of $t=1$. If $\mathcal{CP}$ returns no solution, we keep on incrementing the value of $t$ by one until the minimum length attack vector is returned.

	\subsection{Attack-Resilient Control Execution Sequence Synthesis}
	\label{subsecPatternSyn}
	In this section, we will generate optimal attack-resilient control skip patterns in $2$ steps with respect to the minimum length attack vector $\mathcal{A}$ that the constraint solving problem $\mathcal{CP}$ returns in (\ref{eqMinLenAtt}). \emph{First,} utilizing the condition presented in Theorem~\ref{thCriteria}, we generate a list of $t$ length {\em sub-patterns} that are beneficial in enhancing system's resilience against FDI attacks. In \emph{second} step, we formulate a DP based solution method to compute the final i.e. optimal control execution skip pattern by merging the sub-patterns generated in the first step. The DP based formulation also facilitates generating a list of control execution skip patterns ranked in order of the advantage metric. We now elaborately discuss these two steps.
	\begin{algorithm}[!ht]
	\footnotesize
		\caption{Favourable Sub-pattern Synthesis}
		\label{algPatSyn}
		\begin{algorithmic}[1]
			\Require{State matrices $A$, $B$ and $C$, controller gain $K$, observer gain $L$, sensor limit $\mathcal{Y}$, actuation saturation limit $\mathcal{U}$, detector threshold $Th$, safety envelope $X_s$, initial region $X_0$ of the plant states, minimum execution rate $r_{min}$ of the controller}
			\Ensure{List of favourable sub-patterns $subPatternList$ and their advantage metric $D$}
			\Function{AdvPatSyn}{$A$, $B$, $C$, $K$, $L$, $\mathcal{Y}$, $\mathcal{U}$, $Th$, $X_s$, $X_0$}
			\State $D[i][i]\gets 0\ \forall i\in[1, t]$; $subPatternList\gets null$;\Comment{Initialization}\label{algPatInit}
			\State $\mathcal{A}\gets\Call{CallSolver}{\mathcal{CP},A, B, C, K, L, \mathcal{Y}, \mathcal{U}, Th, X_s, X_0}$;\Comment{Solve $\mathcal{CP}$ in (\ref{eqMinLenAtt})}\label{algPatCallAttSyn}
			\State $t\gets length(\mathcal{A})$;\label{algPatAttLen}
			\State $\triangle r\gets\Call{ResDiffGen}{A, B, C, K, L, X_0, \mathcal{A}, t}$;\Comment{$\triangle r[i] = r^a_i-r_i$ (Eq.~\ref{eqResDeviationPeriodic})}\label{algPatResDiff}
			\For{k=1 to t}\label{algPatForKStarts}
			\For{l=1 to t-k}\label{algPatForLStarts}
			\State $lhs\gets 0$; $rhs\gets 0$; $\rho(k,l)\gets 1^t$;\label{algPatForLInit}
			\For{i=1 to l}\label{algPatForIStarts}
			\If{k>1} $lhs\gets lhs + A^{i-1}B(a^u_{k+l-i}-a^u_{k-1})$;\label{algPatUpdateLHS1}
			\Else\ $lhs\gets lhs + A^{i-1}Ba^u_{k+l-i}$;\Comment{We assume $a^u_0 = 0$}\label{algPatUpdateLHS2}
			\EndIf
			\State $rhs\gets rhs + A^{i-1}L\triangle r[k+l-i+1]$;\label{algPatUpdateRHS}
			\EndFor\label{algPatForIEnds}
			\If{$||lhs||>||rhs||$}\label{algPatIfEStarts}
    			\State $\rho(k,l)\gets 1^k0^{l-k}1^{t-l}$;\label{algPatUpdatePat}
    			\If{$sum(\rho(k,l))\geq r_{min}\times t$} $D[k][k+l]\gets ||lhs||-||rhs||$;\label{algPatIfRate}
    			\Else\ $\rho(k,l)\gets 1^t$;\label{algPatElseRate}
    			\EndIf
			\EndIf\label{algPatIfEEnds}
			\State $subPatternList\gets subPatternList\cup\rho(k,l)$;\label{algPatUpdateSubPatList}
			\EndFor\label{algPatForLEnds}
			\EndFor\label{algPatForKEnds}
			\Return $D$, $subPatternList$;\label{algPatReturn}
			\EndFunction
		\end{algorithmic}
	\end{algorithm}
	\subsubsection{Favourable Sub-pattern synthesis}
	\label{subsubsecSubPatSyn}
	We denote a $t$-length sub-pattern as a binary string of the form  $\rho(k,l) =1^k0^{l-k}1^{t-l}$ where $k>0$ and $l\geq 0$. This implies periodic execution of the controller in first $k$ iterations, followed by execution skips till $l$-th iteration and then $(t-l)$ periodic executions. The control execution following a sub-pattern $\rho(k,l)$ is quantified with an \emph{advantage} value as defined next.
	\vspace{-2mm}
	\begin{Definition}[\bf Advantage value of a sub-pattern]The advantage of control execution that follows the sub-pattern $\rho(k,l)=1^k0^{l-k}1^{t-l}$ of length $t$ over periodic control execution of length $t$ is quantified by the value $||\triangle e^p|| - ||\triangle e^{ap}|| = ||\sum_{i=0}^{l-1}A^iB\triangle a^u_{k+l-1-i}||-||\sum_{i=0}^{l-1} A^iL\triangle r_{k+l-i}||$ (Theorem~\ref{thCriteria}). \hfill$\Box$
	\label{defAdvantage}
	\end{Definition}
	We present a method to synthesize a list \emph{subPatternList} of favourable sub-patterns of the form $\rho(k,l)$ in Algo.~\ref{algPatSyn} which also stores the advantage value of the sub-patterns in a matrix $D$ of size ${t\times t}$. $D[k][l]$ contains the advantage value of $\rho(k,l)$. The inputs to the proposed framework (Fig.~\ref{figToolFlow}) are passed to Algo.~\ref{algPatSyn}.
	In line~\ref{algPatInit}, we initialize the advantage matrix $D$ with $0$ and $subPatternList$ as null. We solve the constraint solving problem $\mathcal{CP}$ in (\ref{eqMinLenAtt}) to generate minimum length stealthy attack $\mathcal{A}$ (line~\ref{algPatCallAttSyn}) and store its length $t$ in line~\ref{algPatAttLen}. 
	By simulating the system's state progression under no attack and under the attack $\mathcal{A}$ in a periodic control execution for $t$ iterations (following the Eq.~\ref{eqLTINoAttack} and \ref{eqLTIUnderAttack}), we compute $\triangle r^a_i$ for all $i\in[1,t]$ and store them in the array $\triangle r$ (line~\ref{algPatResDiff}). The for loop in line~\ref{algPatForKStarts} signifies the possible number of consecutive $1$'s in the sub-pattern $\rho(k,l)$ and the for loop in line~\ref{algPatForLStarts} signifies the number of consecutive $0$'s following the consecutive $1$'s. The LHS and RHS of the criteria $||\sum_{i=0}^{l-1}A^iB\triangle a^u_{k+l-1-i}||>||\sum_{i=0}^{l-1} A^iL\triangle r_{k+l-i}||$ (Theorem~\ref{thCriteria})  are computed in lines~\ref{algPatUpdateLHS1}-\ref{algPatUpdateLHS2} and line~\ref{algPatUpdateRHS} respectively. In line~\ref{algPatIfEStarts}, we check if the LHS is more than the RHS i.e. the difference in estimation error under periodic execution $\triangle e^p$ is more than that of aperiodic execution $\triangle e^{ap}$. If yes, then we generate a new sub-pattern $\rho(k,l)$ by introducing skips from $(k+1)$ to $(k+l)$ in line~\ref{algPatUpdatePat} for generation of new sub-pattern candidate. Next, in line~\ref{algPatIfRate},  we update the advantage value matrix $D$ with $||lhs||-||rhs||$ (which is nothing but $(||\triangle e^p||-||\triangle e^{ap}||)$) if the sub-pattern $\rho(k,l)$ satisfies the minimum execution rate condition. Else, we modify the sub-pattern $\rho(k,l)$ as a periodic one. Next, we include $\rho(k,l)$ (periodic or aperiodic) into $subPatternList$ (line~\ref{algPatUpdateSubPatList}). Finally, the algorithm returns the list of $t$ length sub-patterns of the form $\rho(k,l)$ along with the matrix $D$ which contains the advantage values of those sub-patterns (line~\ref{algPatReturn}).
	The time complexity of Algo.~\ref{algPatSyn} is $O(t^3)$.
	
	\subsubsection{Optimal Attack-resilient Pattern Synthesis}
	\label{subsubsecDP}
		\begin{figure*}[!ht]
	\centering
	\includegraphics[width=\linewidth,clip]{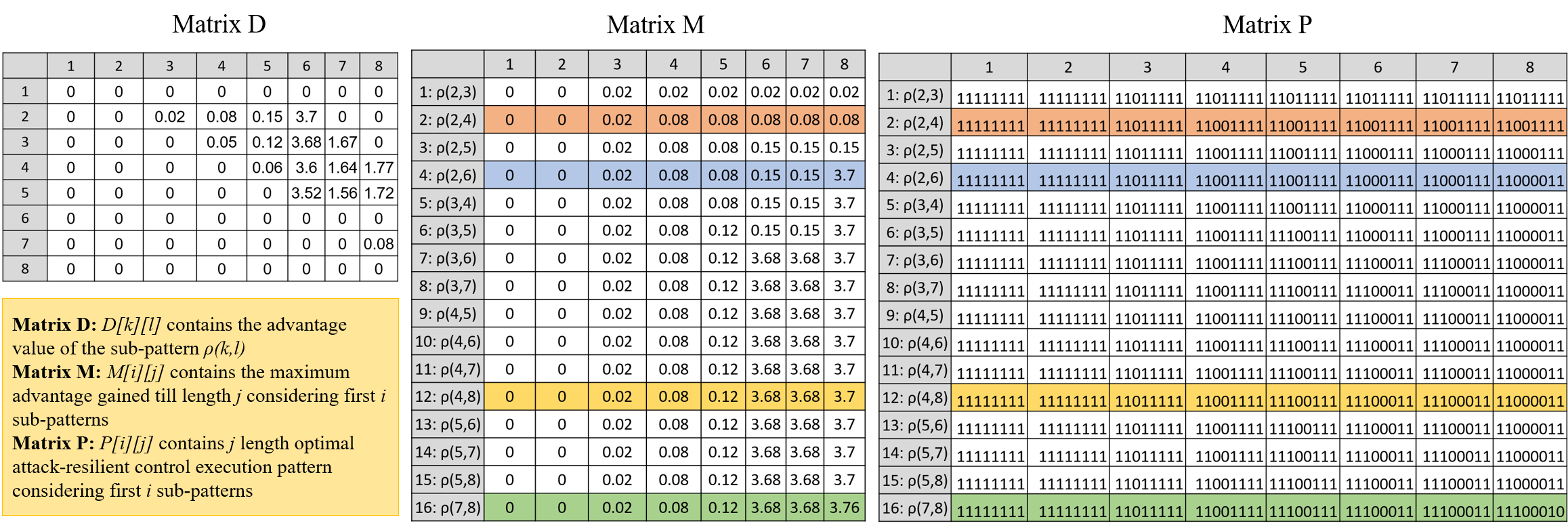}
	\caption{Optimal attack-resilient pattern generation using dynamic programming approach}
	\vspace{-4mm}
	\label{figDP}
	\end{figure*}
	In this section, we formulate a dynamic programming (DP) based  method to synthesize the optimal attack-resilient control execution patterns using the $subPatternList$ and $D$ generated from Algo.~\ref{algPatSyn}. We demonstrate the method with help of the example given in Fig.~\ref{figDP} where the length of the minimum-length attack vector $\mathcal{A}$ is $t=8$ and the minimum rate criteria $r_{min}=0.5$. In the DP formulation, we maintain $2$ matrices: $M_{p\times t}$ and $P_{p\times t}$ where $t$ and $p$ are the length of minimum length attack vector $\mathcal{A}$ and the number of sub-patterns in $subPatternList$. For each sub-pattern $\rho(k,l) \in subPatternList$, $D[k][l]$ has a non-trivial entry. The row indices of $M, P$ and are $subPatternList$ basically lexicographic ordering of such non-trivial $(k,l)$-pairs. For example, since $D[2][3], \cdots, D[2][6]$ are non-trivial, the first four rows of $M, P$ are $1:\rho(2,3),\cdots, 4:\rho(2,6)$ in Fig.~\ref{figDP}.
The maximum advantage value (Def.~\ref{defAdvantage}) that can be achieved by considering skips till $j$-th position in the pattern of length $t$ when only the first $i$ sub-patterns in the $subPatternList$ are taken into consideration, is computed and stored in $M[i][j]$.
The corresponding optimal pattern is stored in $P[i][j]$. 

\par Let the $i$-th sub-pattern be $\rho(k,l) = 1^k0^{l-k}1^{t-l}$. 
For this, let $k = end1(i)$ and $l = end0(i)$ denote the index where the initial 1's and 0's of the sub-pattern $\rho(k,l)$ finish. The rate of control execution for any $t$-length pattern containing a total of $n$ 1s is given by $rate = n/t$. We define merging of two patterns $i$ and $j$ using element wise logical AND operation and denote the merged pattern by  $i\circ j$. 
For example, $D[2][3] = 0.02$ implies the advantage value of $\rho(2,3)$ is $0.02$. With help of this example, we now elaborate how to populate $M$ (Eq.~\ref{EqDPM}) and $P$ using DP memoization process.

	\textbf{case 1} ($i=1$): Consider the $1$-st sub-pattern  in the list \emph{subPatternList} to be used (as per definition) to populate  first row of $M$ and $P$. Since we do not have any favourable sub-pattern with skips up to the length $(end0(1) - 1)$ with non-zero advantage value, we populate $M[1][j] = 0$ and $P[1][j] = 1^t$ for $j<end0(1)$. Let us consider the first row of $M$ and $P$ in the example of Fig.~\ref{figDP}. This is corresponding to the sub-pattern $\rho(2,3)$. Consider the first $end0(1)-1 =2$ columns of $M$. 
	The sub-pattern $\rho({2,3})$ is of the form $1^20^{3-2}1^{8-3}=11011111$, in which there is no prefix up to length $2$ that produces positive advantage value (i.e., $\textbf{[1]}1011111$ or $\textbf{[11]}011111$). 
	Therefore, we update $M[1][1]=M[1][2]=0$ and $P[1][1]=P[1][2]=1^t$. In the first row, for $j\geq end0(1)$, we populate the matrix $M$ with the advantage value of the first sub-pattern and $P$ with the first sub-pattern if the rate of the first sub-pattern up to length $j$ i.e. $1^{end1(1)}0^{end0(1)-end1(1)}$ $1^{j-end0(1)}$=$1^201^{j-3}$ satisfies the rate criteria. Otherwise, $M[i][j]$ and $P[i][j]$ are assigned $0$ and $1^t$ respectively. Let us again consider the example given in Fig.~\ref{figDP}. Since $\rho(2,3)$ satisfies $r_{min}$ up to the length $j\geq end0(1) = 3$, we populate $M[1][j]$ with $D[2][3]$ and $P[1][j]$ with $1^20^{3-2}1^{8-3}=1^201^5\ \forall j\geq3$.
	
	\textbf{case 2} ($i > 1$ and $j<end0(i)$): Now, let us consider the other sub-patterns in  $subPatternList$. There is no prefix up to length $(end0(i)-1)$ in the $i$-th sub-pattern $\mathbf{[1^{end1(i)}0^{end0(i)-end1(i)-1}]}0$ $1^{t-end0(i)}$ that produces positive advantage value. 
	Therefore, for $i>1,j<end0(i)$, we can update $M[i][j]$ with $M[i-1][j]$. 
	Similarly, we assign $P[i][j] = P[i-1][j]$ for $j<end0(i)$. Consider the second row of the matrices $M$ and $P$ in Fig.~\ref{figDP} i.e. the one corresponding to sub-pattern $\rho(2,4) = 1^20^{4-2}1^{8-4} = 11001111$  (highlighted in red). For, $j<end0(2)$ i.e. $j<4$, there exists no prefix that yields  a positive advantage value in the current sub-pattern $\mathbf{[110]}01111$. Therefore, we set $M[2][j]=M[1][j]$ as $M[1][j]$ holds the maximum advantage that can be gained by considering skips until $j$ ($j<4$) positions of a $t=8$ length pattern. 
Similarly $P[2][j]$ is assigned with $P[1][j]$.

{\footnotesize
\begin{align}
	    M[i][j] =\begin{cases} 0 \text{\ if $i=1$ and $j<end0(i)$}\\
	    D[i][j] \text{\ if $i=1$, $j\geq end0(i)\wedge rate(i)\geq r_{min}$}\\
	    0 \text{\ if $i=1$, $j\geq end0(i)$, and $rate(i)<r_{min}$}\\
	    M[i-1][j] \text{\ if $i>1\wedge j<end0(i)$}\\
	    M[i-1][j] \\
	    \quad\text{\ if $i>1\wedge j\geq end0(i)\wedge$}\\
	    \quad\text{$rate(1^{end1(i)}0^{end0(i)-end1(i)}1^{j-end0(i)})<r_{min}$}\\
	    max\{M[i-1][j],D[end1(i)][end0(i)]\} \\
	    \quad\text{\ if $i>1\wedge j\geq end0(i)\wedge $}\\
	    \quad\text{$rate(1^{end1(i)}0^{end0(i)-end1(i)}1^{j-end0(i)})\geq r_{min}\wedge$}\\
	    \quad\text{ $rate(P[i][end1(i)-1]\circ 1^{end1(i)}0^{end0(i)-end1(i)}1^{j-end0(i)})$}\\
	    \quad\quad\quad\quad\quad\quad\quad\quad\quad\quad\quad\quad\quad\quad\quad\quad\quad\quad\quad\quad\quad\quad\quad\quad\text{$<r_{min}$}\\
	    max\{ M[i-1][j], M[i][end1(i)-1] + D[end1(i)][end0(i)]\} \\
	    \quad\text{\ if $i>1\wedge j\geq end0(i)\wedge$,}\\
	    \quad\text{$rate(P[i][end1(i)-1]\circ 1^{end1(i)}0^{end0(i)-end1(i)}1^{j-end0(i)})$}\\
	    \quad\quad\quad\quad\quad\quad\quad\quad\quad\quad\quad\quad\quad\quad\quad\quad\quad\quad\quad\quad\quad\quad\quad\quad\text{$\geq r_{min}$}
        \end{cases}
	    \label{EqDPM}
	\end{align}
}
	\textbf{case 3} ($i > 1$ and $j\geq end0(i)$): 
	We divide this case in $3$ scenarios. 
	\par\textbf{(i)} If the $j$-length prefix of the $i$-th sub-pattern does not satisfy the rate criteria, we set $M[i][j]=M[i-1][j]$ and $P[i][j]=P[i-1][j]$. Consider the  row $i=4$ (highlighted in blue) and column $j=6$ of $M$ in Fig.~\ref{figDP}. The $4$-th row corresponds to the sub-pattern $\rho(2,6)=1^20^{6-2}=11000011$. In $M[4][6]$, we want to have the maximum advantage value that can be achieved by considering skips until $6$-th position of the $8$-length pattern. And, we can compute $M[4][6]$ by considering only the first $4$ sub-patterns i.e. $\rho(2,3)$, $\rho(2,4)$, $\rho(2,5)$ and $\rho(2,6)$. However, the $j=6$ length prefix of $\textbf{[110000]}11$ does not satisfy $r_{min}=0.5$. So, we set $M[4][6] = M[3][6]$ and $P[4][6] = P[3][6]$. \par\textbf{(ii)} If the $j$-length prefix of the $i$-th sub-pattern satisfies the rate criteria, we can consider merging the $i$-th sub-pattern with the most favourable {\em non-overlapping}  sub-patterns. Two sub-patterns are \emph{non-overlapping} if they do not have $0$'s at same position. Therefore, the candidate patterns which can be merged with $i$-th sub-pattern $1^{end1(i)}0^{end0(i)-end1(i)-1}01^{t-end0(i)}$ must have $0$'s before their $end1(i)$-th position. As per the construction of the $P$ matrix, the most favourable candidate sub-pattern for merging with $i$-the sub-pattern is stored in $P[i][end1(i)-1]$. However, the $i$-th sub-pattern can be merged with $P[i][end1(i)-1]$ if the $j$-length prefix of the merged pattern satisfies minimum rate $r_{min}$ criteria even the if they are non-overlapping. If the rate condition on the merged pattern is not satisfied, we set $M[i][j] = max\{M[i-1][j], D[end1(i)][end0(i)]\}$ and accordingly populate $P[i][j]$. Consider the  row $i=12$ (highlighted in yellow) and column $j=8$ of $M$ in Fig.~\ref{figDP}. The $12$-th row corresponds to the sub-pattern $\rho(4,8)=1^40^{8-4}1^0=11110000$. In $M[12][8]$, we want to have the maximum advantage value that can be achieved by considering skips until last position of $8$-length pattern. And, we can compute $M[12][8]$ by considering only the first $12$ sub-patterns i.e. $\rho(2,3)$, $\rho(2,4)$, $\cdots$, $\rho(4,8)$. The pattern $\rho(4,8)$ satisfies the rate condition, and is also mergable with non-overlapping sub-pattern in $P[12][3] = 1101111$. However, the merged pattern $P[12][3]\circ\rho(4,8) = 11011111\circ11110000 = 11010000$ does not satisfy $r_{min} = 0.5$. Thus, we set $M[12][8] = max\{M[11][8], D[4][8]\} = M[11][8]$, and accordingly set $P[12][8] = P[11][8] = 11000011$. 
	\par\textbf{(iii)} Finally, consider that the merged pattern of $i$-th sub-pattern and the pattern in $P[i][end1(i)-1]$ satisfies the rate condition. Then, we check if we can yield better advantage after merging. If so, we set $M[i][j] = M[i][end1(i)-1]+D[end1(i)][end0(i)]$ and $P[i][j] = P[i][end1(i)-1]\circ 1^{end1(i)}0^{end0(i)-end1(i)}1^{t-end0(i)}$. Otherwise, we assign $M[i][j] = M[i-1][j]$ and $P[i][j] = P[i-1][j]$. For example, consider the case corresponding to the $16$-th row i.e. the last sub-pattern $\rho(7,8)$ and the maximum length i.e. $8$ in Fig.~\ref{figDP} (highlighted in green). If $\rho(7,8)$ is merged with $P[16][6]$ i.e. $1^30^01^2$, we get $1^30^310$ which satisfies the $r_{min}=0.5$. This merging gives an advantage value of $M[16][6] + D[7][8] = 3.76$ which  is more than the maximum advantage value computed (i.e. $M[15][8]$)  before considering $\rho(7,8)$ for $8$ length patterns. Therefore, we populate $M[16][8]$ with $3.76$ and $P[16][8]$ with $1^30^310$. 
	\par The last column of $P$ matrix i.e. $P[i][t]\forall i\in[1,p]$ gives the list of attack-resilient control execution patterns of length $t$ ranked (from least beneficial to most beneficial) with respect to the advantage values.Thus, by construction, the most attack-resilient $t$ length optimal control execution pattern is stored in $P[p][t]$. We can see in the example of Fig.~\ref{figDP} that $M[16][8]$ has the maximum advantage value with the corresponding pattern  stored in $P[16][8]$. The time complexity of this DP based solution method is $O(pt)$. 
	
\section{Experimental Results}
\label{secResult}
For evaluation of our framework, we consider several safety-critical CPS  benchmarks. The system descriptions are given as the input to our tool along with their initial region, performance and safety criteria as mentioned in Fig.~\ref{figToolFlow}. 
The framework is built using Matlab and is shared in a public repository \footnote{https://anonymous.4open.science/r/OptimalPatternSynthesis-F8E2/}. Our experiments are run on an 8-core 7-th gen intel i7 CPU with 16 GB of RAM. 

\begin{table}[!ht]
\small{
\vspace{0.5mm}
\caption{\small Resilient Control Sequences for Automotive Benchmarks}
\label{tabResult}
\scalebox{0.785}{
\begin{tabular}{|l|c|c|c|l|c|c|}
\hline
Systems &
  \begin{tabular}[c]{@{}c@{}}Dime- \\ -nsion\end{tabular} &
  $r_{min}$ &
  \begin{tabular}[c]{@{}c@{}}Minimum \\ length of \\ Attack\end{tabular} &
  \multicolumn{1}{c|}{\begin{tabular}[c]{@{}c@{}}Pattern\\ Synthesis\\ Time (s)\end{tabular}} &
  \begin{tabular}[c]{@{}c@{}}Control\\ Execution\\ Sequences\end{tabular} &
  Advantages \\ \hline
\multirow{4}{*}{\begin{tabular}[c]{@{}l@{}}Trajectory\\  Tracking\\  Control~\cite{adhikary2020skip}\end{tabular}} &
  \multirow{4}{*}{2} &
  \multirow{4}{*}{0.51} &
  \multirow{4}{*}{13} &
  \multirow{4}{*}{\begin{tabular}[c]{@{}l@{}}0.55s\\ {[}Total:\\ 21.55s{]}\end{tabular}} &
  $10^61^6$ &
  13.58 \\ \cline{6-7} 
 &  &  &  &  & $10^51^7$       & 9.86    \\ \cline{6-7} 
 &  &  &  &  & $10^41^8$       & 6.14    \\ \cline{6-7} 
 &  &  &  &  & $10^31^9$       & 2.05    \\ \hline
ESP~\cite{adhikary2020skip} &
  2 &
  0.45 &
  3 &
  \begin{tabular}[c]{@{}l@{}}0.054s\\ {[}Total: 4.18s{]}\end{tabular} &
  101 &
  2.62 \\ \hline
\multirow{4}{*}{\begin{tabular}[c]{@{}l@{}}Fuel\\ Injection\\~\cite{wei2009modeling}\end{tabular}} &
  \multirow{4}{*}{3} &
  \multirow{4}{*}{0.5} &
  \multirow{4}{*}{8} &
  \multirow{4}{*}{\begin{tabular}[c]{@{}l@{}}0.38s\\ {[}Total:\\ 13.26s{]}\end{tabular}} &
  $10^31^30$ &
  7.03 \\ \cline{6-7} 
 &  &  &  &  & $10^21^30^2$    & 6.49    \\ \cline{6-7} 
 &  &  &  &  & $1^50^3$        & 4.22    \\ \cline{6-7} 
 &  &  &  &  & $1^40^4$        & 3.86    \\ \hline
\multirow{3}{*}{\begin{tabular}[c]{@{}l@{}}Suspension\\  Control~\cite{roy2016multi}\end{tabular}} &
  \multirow{3}{*}{4} &
  \multirow{3}{*}{0.52} &
  \multirow{3}{*}{4} &
  \multirow{3}{*}{\begin{tabular}[c]{@{}l@{}}0.12s\\ {[}Total:\\ 7.08s{]}\end{tabular}} &
  $1^20^2$ &
  2274.73 \\ \cline{6-7} 
 &  &  &  &  & $10^21$         & 2096.29 \\ \cline{6-7} 
 &  &  &  &  & $101^2$         & 434.14  \\ \hline
\multirow{4}{*}{\begin{tabular}[c]{@{}l@{}}Four-Car \\ Platoon~\cite{schurmann2017optimal}\end{tabular}} &
  \multirow{4}{*}{8} &
  \multirow{4}{*}{0.5} &
  \multirow{4}{*}{25} &
  \multirow{4}{*}{\begin{tabular}[c]{@{}l@{}}3.56s\\ {[}Total:\\ 272.62s{]}\end{tabular}} &
  $10^{12}1^{12}$ &
  4.64 \\ \cline{6-7} 
 &  &  &  &  & $10^{11}1^{13}$ & 4.01    \\ \cline{6-7} 
 &  &  &  &  & $10^{10}1^{14}$ & 3.44    \\ \cline{6-7} 
 &  &  &  &  & $10^{9}1^{15}$  & 2.92    \\ \hline
\end{tabular}
}}
\end{table}

\begin{figure}[!hb]
	\begin{subfigure}[b]{0.5\columnwidth}
		\includegraphics[width=\textwidth,keepaspectratio,clip]{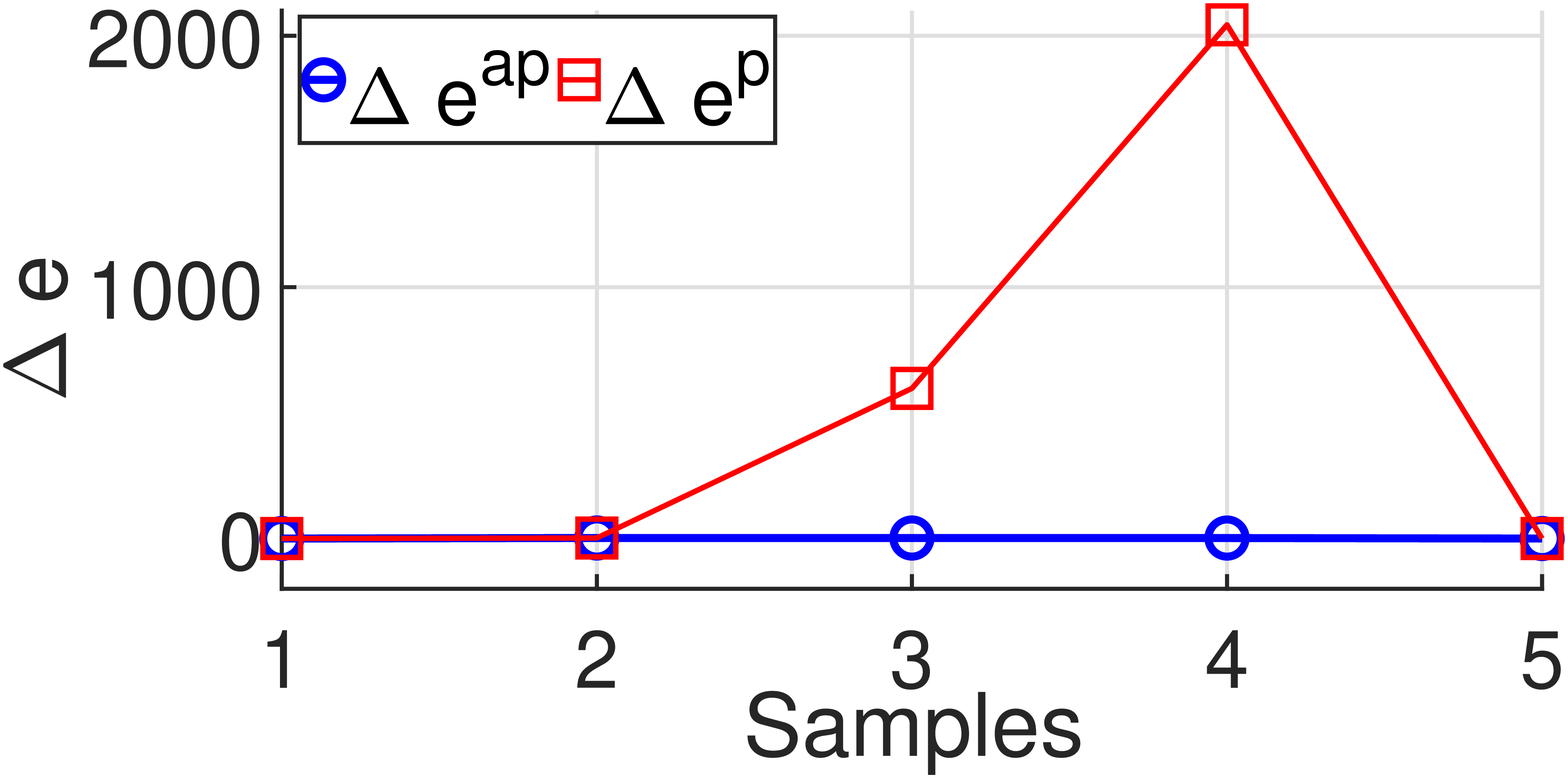}
		\caption{Comparison Between $\Delta e^p$ and $\Delta e^{ap}$}
		\label{figSUspConEst}
	\end{subfigure}
	\hfill
	\begin{subfigure}[b]{0.49\columnwidth}
		\includegraphics[width=\textwidth,keepaspectratio,clip]{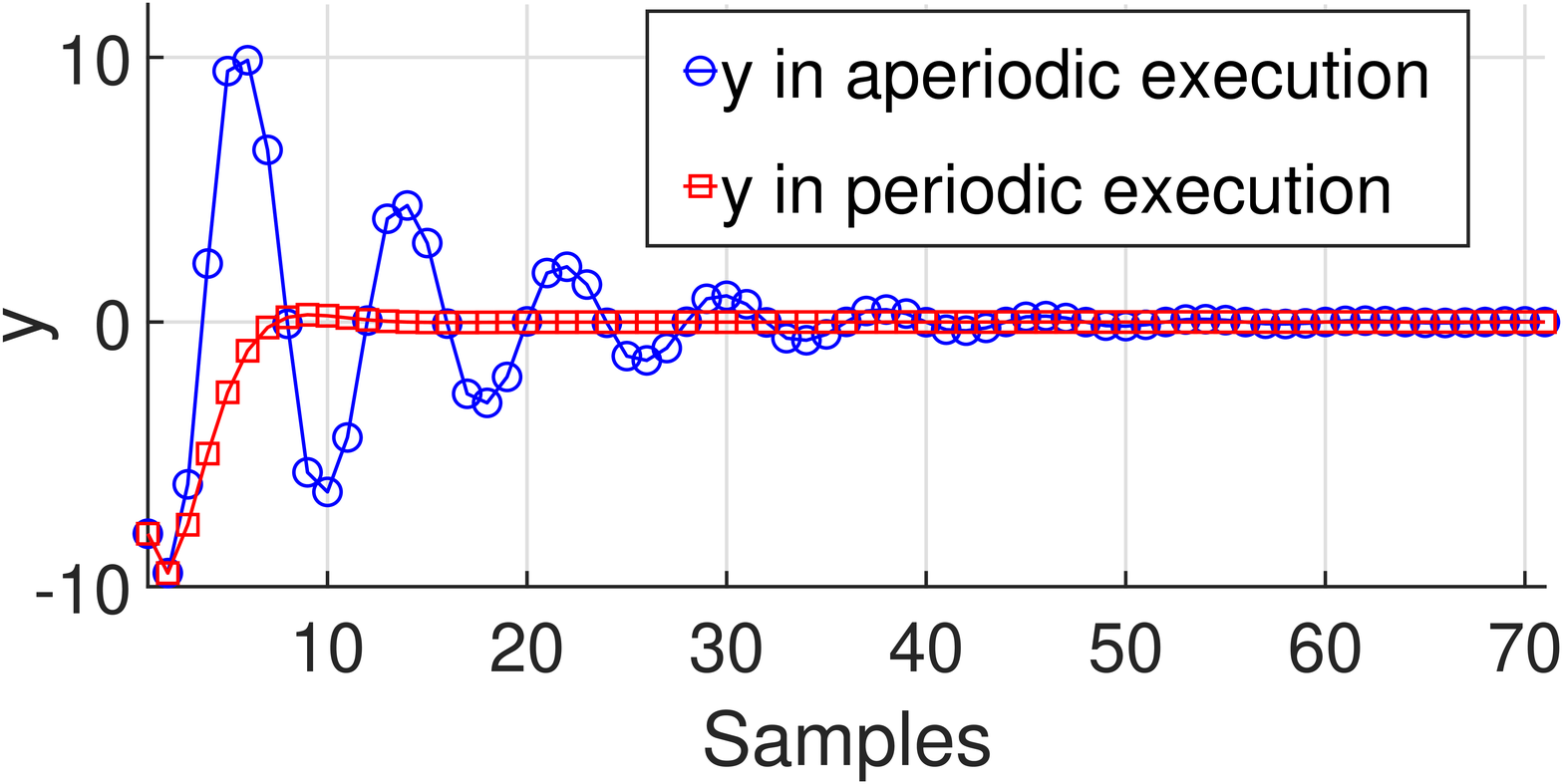}
		\caption{Comparison Between periodic $y$ and aperiodic $y$ under no FDI}
		\label{figSuspConOp}
	\end{subfigure}
\caption{Effect of Derived Aperiodic Control Execution Sequence $(1001)^{\omega}$ on Suspension Control System under FDI}
\label{figSuspCon}
\end{figure}

In Tab.~\ref{tabResult} we demonstrate  attack-resilient control execution sequences synthesized for control systems (provided with corresponding references in the 1st column) with different dimensions (provided in the 2nd column of the table) in order to verify the scalability of our approach. The synthesized patterns satisfy the minimum execution rate $r_{min}$ (3rd column) and their length is considered same as the minimum-length attack (4th column) discovered using Eq. 
\ref{eqMinLenAtt}-\ref{eqAttSafety}. The 6th and 7th column provides the  list of synthesized control execution patterns  ranked in descending order w.r.t. resilience and their advantage values. In the 5-th column we provide the runtime of the pattern synthesis methodology (along with the runtime of the overall methodology in braces). 
The system descriptions along with  safety and performance criteria are taken from ~\cite{roy2016multi,suspCTMS}. Further, in Figure ~\ref{figSuspCon}, we demonstrate the resilience and performance of the best (w.r.t the advantage value) synthesized control execution sequence for a suspension control system~\cite{roy2016multi}.
  Our framework generates a minimum-length attack for this system which can make the system unsafe within 4 sampling iterations. The generated best possible attack-resilient control sequence is $1001$ (the 1st sub-row in the 6-th column of the 4-th row of the Tab.~\ref{tabResult}). The blue plot with circle marker denotes $\triangle e^{ap}$ while following the control execution pattern $1001$ and the red plot with square marker denotes $\triangle e^{p}$ while following periodic control execution. As we can clearly observe in Fig.~\ref{figSUspConEst}, under the 4-length attack $\triangle e^{ap}$ is significantly less than $\triangle e^{p}$. Due to this, the estimation error induced by the false data is significantly reduced by repeating the control execution skips at every 2-nd and 3-rd position of a 4 length execution/skip pattern. Fig.~\ref{figSuspConOp} showcases the performance of the system under the aperiodic control sequence $1001$ (in blue and circled plot) and the periodic control execution (in red and squared plot) without any FDI attack. As per the design criteria, the system must settle within 3 seconds. We can see system output (position of the car in meters) under the periodic control execution settles much quickly compared to the aperiodic execution. However, since the aperiodic control sequences synthesized using our framework always follows the $r_{min}$, even under the aperiodic execution the system output settles within 2.4seconds (i.e. 60  sampling periods each of 0.04 sec). This successfully validates that the attack-resilient control sequences generated using our framework preserve  system performance while turning out to be beneficial in terms of reducing the damage caused by an FDI. 
 
  \begin{figure}[]
\begin{subfigure}[b]{0.5\columnwidth}
	\includegraphics[width=\textwidth,keepaspectratio,clip]{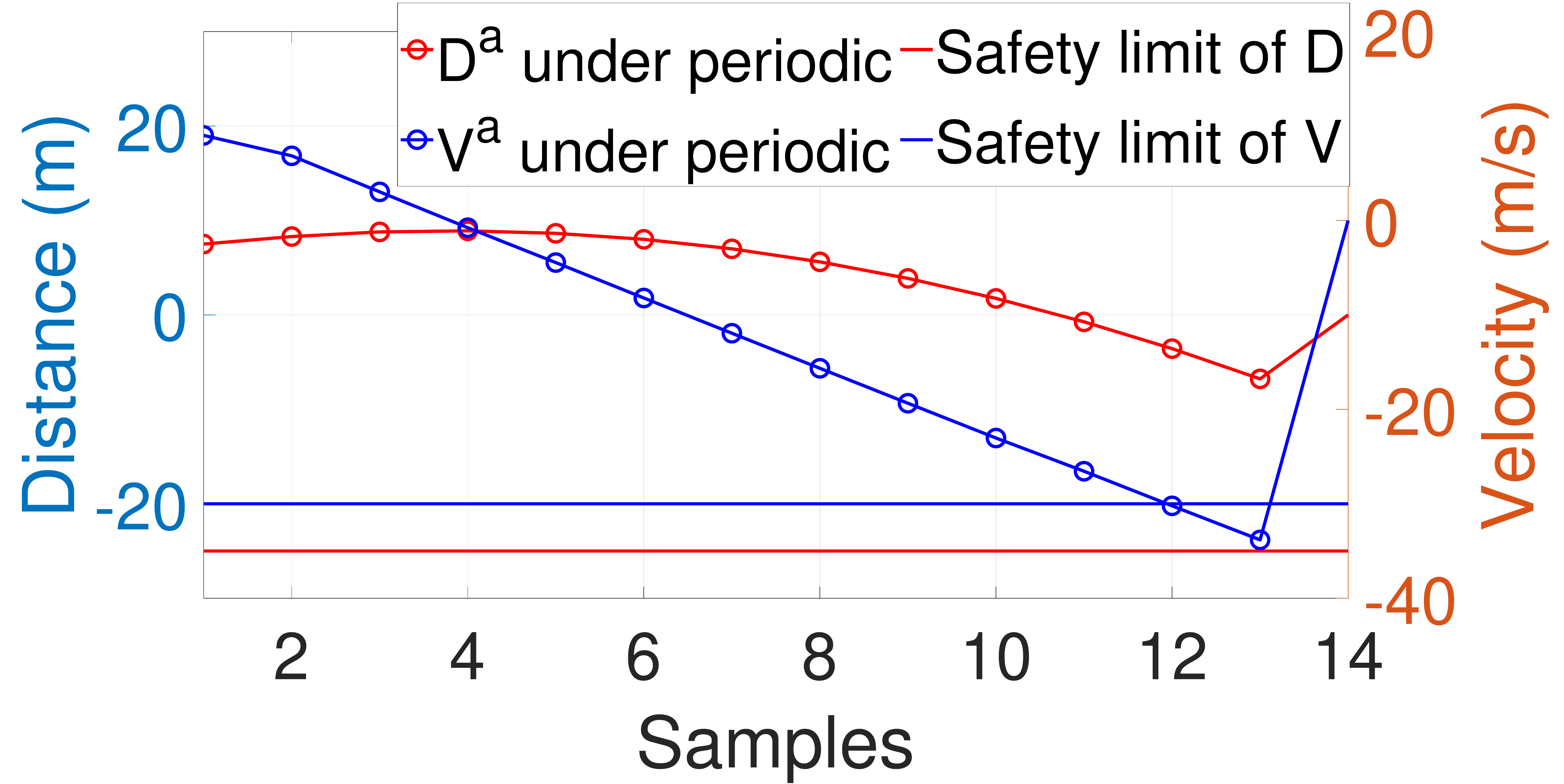}
	\caption{TTC under minimum-length attack in periodic execution}
	\label{figTTCPeriodic}
\end{subfigure}
\hfill
\begin{subfigure}[b]{0.49\columnwidth}
	\includegraphics[width=\textwidth,keepaspectratio,clip]{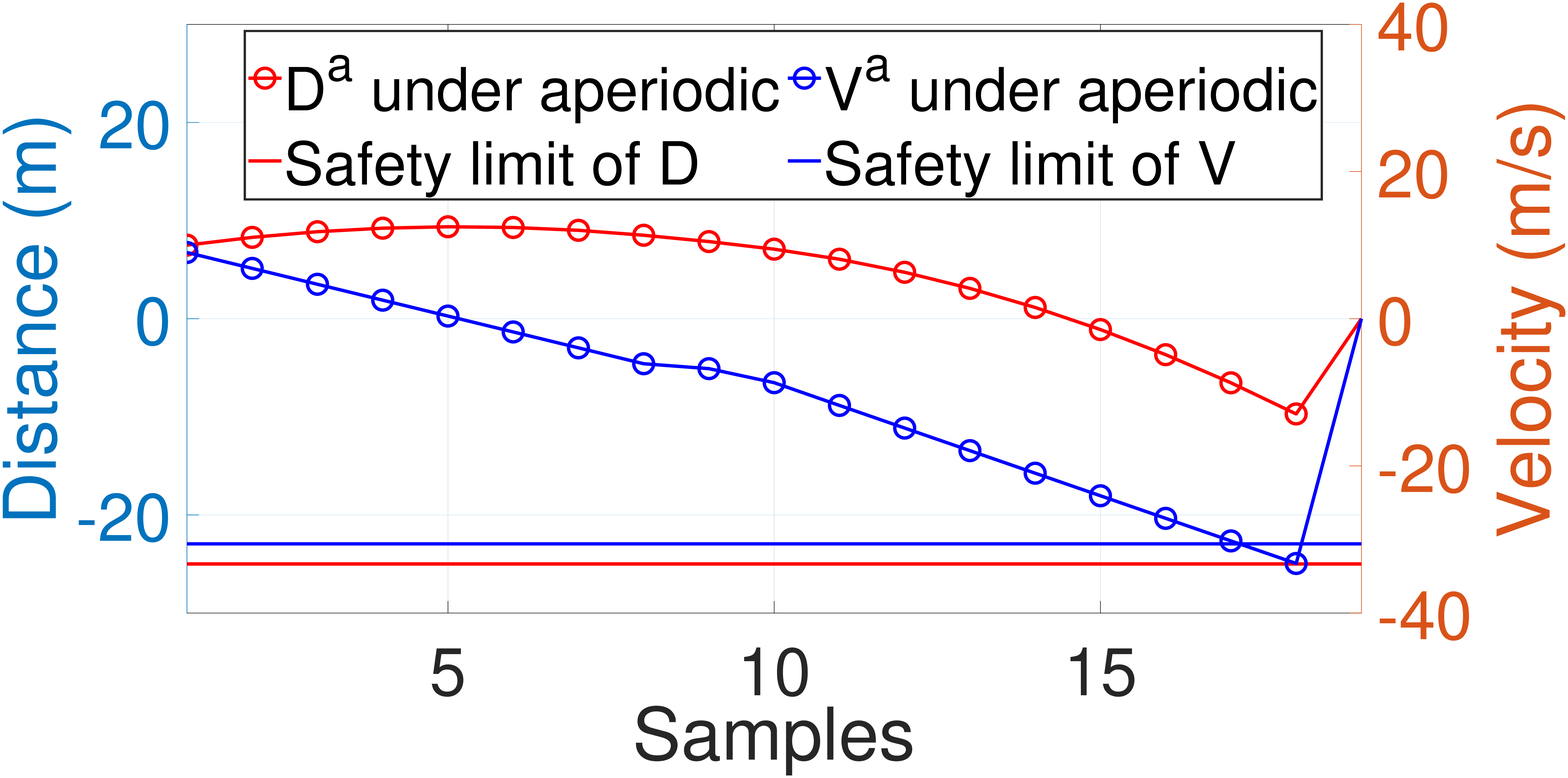}
	\caption{TTC under minimum-length attack under $10^61^6$}
	\vspace{-2mm}
	\label{figTTCAPeriodic}
\end{subfigure}
\caption{Effect of aperiodic control execution sequence on trajectory tracking control system under FDI attack}
\label{figTTC}
\end{figure}
\subsubsection*{\textbf{A Use case:}  
Lightweight Security Design Utilizing Control Execution Sequence for Automotive CPS} Automotive CPSs are safety-critical but often  resource-constrained. 
Thus we cannot afford to secure the closed loop communications in every iteration. The state-of-the-art technique to handle this resource-aware security design is to activate the cryptographic measures intermittently~\cite{jovanov2019relaxing}. For this we can utilize the aperiodic control executions, synthesized using our framework, to make the system resilient enough against a minimum-length attack sequence while the cryptographic encryption is not active. As we have seen earlier, (refer Sec.~\ref{subsecAttackGeneration}) an attack vector generated using our framework commits maximum effort to make the system states unsafe without being detected in minimum possible time. But if we choose to run the closed loop control execution following the synthesized optimally attack-resilient control sequence, the attack effect (the estimation error difference $\Delta e$) can be kept in check for a longer duration. This enables the system to behave in a more resilient fashion under FDI attacks resulting to a less frequent activation of the cryptographic method. Let us visualize such a scenario. As we can see in Fig.~\ref{figTTCPeriodic}, under the periodic control execution sequence, outputs of the TTC i.e. distance from the desired trajectory ($D$ in meters, the blue plots) and velocity of the vehicle ($V$ in m/s, the red plots) goes beyond the safety boundaries at 13-th sampling iteration. But while following the best aperiodic execution pattern synthesized using our framework for TTC i.e. $10^61^6$, the system does not become unsafe at 13-th iteration. Rather as we can see in Fig.~\ref{figTTCAPeriodic}, the generated minimum-length attack for $10^61^6$ is of 18 length, i.e. the attack  makes the system unsafe at 18-th sampling iteration when operated with $10^61^6$.  This simply suggests that the system under the synthesized aperiodic execution can promise more resilience against attack and can reduce the activation of the cryptographic method by $\sim 21\%$ (activation of crypto can be delayed from 14 sampling iterations in case of periodic executions to 17 in case of aperiodic executions). This motivates the fact such optimally attack-resilient control sequences can be useful in resource-aware CPS co-designs. 
	\section{Conclusion}
	\label{secConclusion}
	In this work, 1) we establish analytical  conditions under which occasional control skips improve the system resilience w.r.t. FDI attacks, and 2) provide an associated CAD  framework for generating such skip sequences. Extending the constraint  solving problem formulation (Eq. 10-16) in our methodology with a  counter-example guided loop, it is possible to generate multiple attack vectors of minimum length and beyond the minimum length. For all such cases, applying the  control execution sequence generation method provides a library of robust control schedules which can be deployed in a CPS. 
	Creating a statistical foundation for choosing among such sequences given the probability  distribution of attack  vectors is considered as future work. Also, as the use case suggests,  using attack-resilient control executions for relaxing real-time resource constraints  in a methodical manner can be another future extension.


\bibliographystyle{ACM-Reference-Format}
\bibliography{reference}

\end{document}